\documentclass[11pt]{article}
\usepackage{geometry}                
\usepackage{graphicx}
\usepackage{amssymb,amsmath,bbm,natbib}
\usepackage{epstopdf}
\usepackage{color}
\graphicspath{{./Figs/}}

\newcommand{\X}{\boldsymbol X}

\newcommand{\E}{\mathbb E}

\DeclareMathOperator*{\argmax}{arg\,max}

\title{A Change-Point Model for Detecting Heterogeneity in Ordered Survival Responses}
\author{O. Bouaziz and G. Nuel}

\begin{document}
\maketitle

\begin{abstract}
In this article we suggest a new statistical approach considering survival heterogeneity as a breakpoint model in an ordered sequence of time to event variables. The survival responses need to be ordered according to a numerical covariate. Our estimation method will aim at detecting heterogeneity that could arise through the ordering covariate. We formally introduce our model as a constrained Hidden Markov Model (HMM) where the hidden states are the unknown segmentation (breakpoint locations) and the observed states are the survival responses. We derive an efficient Expectation-Maximization (EM) framework for maximizing the likelihood of this model for a wide range of baseline hazard forms (parametrics or nonparametric). The posterior distribution of the breakpoints is also derived and the selection of the number of segments using penalized likelihood criterion is discussed. The performance of our survival breakpoint model is finally illustrated on a diabetes dataset where the observed survival times are ordered according to the calendar time of disease onset.

\end{abstract}

\section{Introduction}
\label{intro}

In survival analysis it is quite common that heterogeneity between patients results in various survival response distributions. This heterogeneity can be controlled through known covariates (such as date of birth, age at diagnosis, gender, treatment, co-exposure, BMI, etc.) using regression-type models such as the Cox proportional hazard model (\cite{cox}) and by performing stratified analyses or by incorporating a random effect in a frailty model (see among many other authors~\cite{clayton78},~\cite{hougaard95},~\cite{therneau00} and \cite{ripatti02}). 
Other types of heterogeneous dataset arise when the incidence rate changes over the calendar time in a cohort study and specific models like age-period-cohort have been extensively studied to take into account this kind of heterogeneity (see~\cite{Yang} for instance). While theses models have proved to be most useful, it is however likely that unaccounted latent heterogeneity remains in the survival signal. This might be due for example to an unknown interaction between a treatment and some exposure, or to some unaccounted heterogeneity of the disease itself (for example an unknown cancer sub-type).

Fitting heterogeneous survival models such as frailty models or cure models (see for instance~\cite{farewell82} and~\cite{sy00}) is a challenging task which often requires specifying parametric incidence rates in order to ensure identifiability. When considering nonparametric hazard rates the task is even more challenging and usually requires additional constraints. Quoting~\cite{sy00} in the cure model context, ``{\it{by leaving the conditional baseline survival function arbitrary, a condition close to nonidentifiability can occur, which causes estimation problems}}'' and they further mention that this issue is overcome by requiring the additional constraint that the conditional survival function is set to zero beyond the last event time. In the frailty context, \cite{rondeau07} overcome the nonidentifiability issue by using a smoothing approach : the authors consider spline functions for the estimation of the baseline hazard and a penalized likelihood estimation method is implemented in order to estimate the regression parameters while controlling the smoothness of the baseline hazard.

In the present work, we suggest a new approach considering survival heterogeneity as a breakpoint model in an ordered sequence of survival responses. The survival responses might be ordered according to any numerical covariate (ties are possible) like age at diagnosis, BMI, etc. The basic idea being that heterogeneity will be detected as soon as it is associated with the chosen covariate. For instance, age at diagnosis might be associated with a higher chance to receive a new treatment or BMI might be associated with a specific exposure. 
From a statistical point of view we consider this situation as a change-point model where abrupt changes occur in terms of baseline hazard rates and/or in terms of proportional factors. In such a model, we aim at two objectives: first we want to estimate the hazard rates and the proportional factors in each homogenous region through a Cox model considering parametric baseline hazards or a nonparametric baseline hazard. Secondly, we want to accurately provide the number and location of the breakpoints. Recently a constrained Hidden Markov Model (HMM)  method was suggested in the context of breakpoint analysis \citep[see][]{Nuel}. This method allows to perform a full change-point analysis in a segment-based model (one parameter by segment) providing linear EM estimates of the parameter and a full specification of the posterior distribution of change points. In this paper we adapt this method to the context of survival analysis with hazard rate estimates, where the estimation is performed through the EM algorithm \citep[see][]{Dempster1977} to provide update of the estimates and the posterior distribution at each iteration step. 

In~\cite{ABGK}, the authors studied a dataset on nephropathy for diabetics (introduced in Example I.3.11 of their book) using a multi-state model, where each transition intensity models was adjusted with respect to the calendar time of disease onset (see Table VII.2.1 page 520 of their book). The authors concluded that ``\emph{it is seen that all intensities decrease with $t_0$ (the calendar year of onset of diabetes), indicating a general medical improvement over time}''. We will illustrate our method on this dataset, where the event times will be ordered with respect to the calendar time of disease onset and our model will aim to detect heterogeneity on the survival distribution of the patients with respect to the calendar time of disease onset.

 In Section~\ref{sec:model}, the Cox breakpoint model and the corresponding conditional likelihood are presented. In Section~\ref{EMalgo}, the EM algorithm is introduced as an iterated method to perform estimation in this context. It is shown that the E step can be seen as a weighted likelihood where the weights correspond to the posterior probability of each individual to be in each segment given the data and the previous update of the model parameter. In Section~\ref{sec:posterior}, computation of the weights is derived. In Section~\ref{sec:maxim}, maximisation of the log-likelihood for a fixed weight is discussed. Three parametric baseline hazards (exponential, Weibull or piecewise constant) and the nonparametric baseline are studied in our model and their expressions are recalled in the Supporting Material. 
 Section~\ref{sec:global} gives a summary of the implementation of the proposed algorithm along with some discussions on the calibration of the algorithm parameters. A simulation study is presented in Section~\ref{sec:simu}. Section~\ref{sec:robust} discusses the ability of our BIC criterion to accurately find the correct number of breakpoints in the data and a real data analysis on survival of diabetic patients is studied in Section~\ref{data}. Finally, Section~\ref{sec:conclusion} concludes this article with some general comments on the proposed methods.

\section{Model and estimation procedure}

\subsection{The breakpoint model}\label{sec:model}

Let $T^*$ represent the survival time of interest associated with its counting process $N^*(t)=I(T^*\leq t)$ and its at risk process $Y^*(t)=I(T^*\geq t)$ for $t\geq 0$. Let $\X$ represent a $p$-dimensional covariate row vector. In practice, $T^*$ might be censored by a random variable $C$ so that we observe $(T=T^*\wedge C, \Delta=I(T^*\leq C),\X)$. Introduce the observed counting and at risk processes denoted respectively by $N(t)=I(T\leq t, \Delta=1)$ and $Y(t)=I(T\geq t)$ and let $\tau$ be the endpoint of the study. The data consist of $n$ independent replications $(T_i, \Delta_i,\X_i)_{i=1,\ldots,n}$ associated with their counting process $N_i(t)$ and at risk process $Y_i(t)$, for $t\in[0,\tau]$.


The cohort effect is modelized through the latent random variable $R$ and its $n$ i.i.d. replications $R_1,R_2,\ldots , R_n$ which represent an unobserved segment index associated to each individual. We suppose that the population is composed of $K$ segments such that for $i=1,\ldots, n$, $R_i\in \{1,2,\ldots, K \}$. Without loss of generality, we also assume that the $R_is$ are ordered. For example, if the population is a mixture of three subpopulations such that we have $n=10$ and two breakpoints occurring after positions $3$ and $7$ then $R_{1:10}=1112222333$.


The goal of this paper is to study a hazard Cox model stratified with respect to the segment index. This model is defined in the following way:
\begin{align}\label{model}
\E[dN^*(t)|Y^*(t),\X,R]=Y^*(t)\sum_{k=1}^K\lambda_k(t)\exp(\boldsymbol{X} \boldsymbol{\beta}_k)I(R=k)dt,
\end{align}
where the $\lambda_k$ represent unknown baseline hazard functions and the $\boldsymbol{\beta}_k$ unknown regression parameters associated to each segment index. Let $\Lambda_k(t)=\int_0^t \lambda_k(s)ds$ represents the cumulative baseline hazard function of the $k$th segment index. We denote by $\boldsymbol{\theta}=(\Lambda_1,\ldots,\Lambda_K, \boldsymbol{\beta}_1, \ldots, \boldsymbol{\beta}_K)$ the model parameter we aim to estimate. 
Note that if the $R_is$ were observed and if $\boldsymbol\beta_1=\cdots=\boldsymbol\beta_K$, this model would reduce to the classical stratified Cox model \citep[see for instance][page 190]{martinussenscheikebook}.

In order to make inference on the model parameter we will assume that the endpoint $\tau$ is defined such that, for all $t$ in $[0,\tau]$,
$\mathbb P(T>t)>0.$ We will also suppose that the censoring variable is independent of the event time conditionally on $\X$ and $R$. Under this independent censoring assumption, our model defined by Equation~\eqref{model} is still verified if we replace the processes $N^*(t)$ and $Y^*(t)$ by their observed counterpart, namely $N(t)$ and $Y(t)$.
The contribution of the $i$th individual to the likelihood conditionally on its (unobserved) segment index being equal to $k$ is represented by
\begin{equation*}
e_i(k;\boldsymbol{\theta})=\mathbb{P} (T_i,\Delta_i,\X_i | R_i=k; \boldsymbol{\theta}). 
\end{equation*}
 From standard arguments on likelihood constructions in the context of survival analysis \citep[see for instance][]{ABGK}, we have under independent and non informative censoring: 
\begin{equation}\label{likelihood}
\log e_i(k;\boldsymbol{\theta}) =\int_0^{\tau}\left\{\log\big(\lambda_k(t)\big)+\X_i\boldsymbol\beta_k\right\}dN_i(t)\\
-\int_0^{\tau}Y_i(t)\lambda_k(t)\exp(\X_i\boldsymbol\beta_k)dt, 
\end{equation}
where the equality holds true up to a constant that does not depend on the model parameter $\boldsymbol{\theta}$. Since the segment indexes are not observed, the likelihood of our model 
cannot be directly computed. To overcome this problem, an Expectation-Maximization (EM) algorithm procedure is developed in the next section. 
%
%
\subsection{The EM algorithm}\label{EMalgo}

By considering the segmentation $R_{1:n}=R_{1},\ldots,R_n$ as a latent variable, the EM-algorithm \citep[see][]{Dempster1977} consists of performing alternatively until convergence the following two-steps.
\begin{description}
\item{\bf Expectation Step:} compute the conditional expected log-likelihood
\begin{equation*}
Q(\boldsymbol{\theta} | \boldsymbol{\theta}_\text{old}) = \int_{R_{1:n}} \mathbb{P}(R_{1:n} | \text{data}; \boldsymbol \theta_\text{old} ) 
\log \mathbb P(R_{1:n},\text{data}; \boldsymbol \theta ) dR_{1:n}
\end{equation*}
where $\boldsymbol \theta_\text{old}$ denote the previous value of the parameter and $\text{data}=(T_{1:n},\Delta_{1:n},\X_{1:n})$.\\
\item{\bf Maximization Step:} update parameter with 
\begin{equation}\label{EM}
\boldsymbol{\hat \theta}= \underset{\boldsymbol \theta } \argmax \,Q(\boldsymbol{\theta} | \boldsymbol{\theta}_\text{old}).
\end{equation}
\end{description}

Assuming that the prior segmentation distribution $\mathbb{P}(R_{1:n}; \boldsymbol \theta)$ does not depend on $\boldsymbol \theta$, we easily get (for details see the Supporting Material, Section 1:
\begin{equation}\label{eq:Q}
Q(\boldsymbol{\theta} | \boldsymbol{\theta}_\text{old})=
\sum_{i=1}^n \sum_{k=1}^K w_i(k; \boldsymbol{\theta}_\text{old}) \log e_i(k;\boldsymbol{\theta}),
\end{equation}
where for any $i \in \{1,\ldots,n\}$, $k \in \{1,\ldots,K\}$ and $\boldsymbol \theta$ we define:
\begin{equation*}
w_i(k; \boldsymbol{\theta})=\mathbb{P}(R_{i}=k | \text{data}; \boldsymbol \theta ).
\end{equation*}

Our EM algorithm hence alternates two steps. First, the E-Step which consists of computing the weights $w_i(k; \boldsymbol{\theta}_\text{old})$. This is done in Section~\ref{sec:posterior} using a constrained Hidden Markov Model (HMM). Then for the M-Step, Equation~\eqref{EM} needs to be solved. This is done in Section~\ref{sec:maxim} using the weighted log-likelihood expression given by Equation~(\ref{eq:Q}).


\section{Computation of the posterior segment distribution}\label{sec:posterior}

As suggested in \citet{Nuel}, the posterior segmentation distribution can be obtained using the constrained HMM. For completeness, we give all the necessary information to implement this constrained HMM. The basic idea consists of modeling the segmentation variable $R_{1:n}$ using a Markov chain over $\{1,\ldots,K,K+1\}$ where $K+1$ is an absorbing (technical junk) state. The segmentation always start with $R_1=1$ and its transition matrix $\mathbb{P}(R_i | R_{i-1})$ is given by the following matrix (in the particular case where $K=4$):
\begin{equation*}
\left(\begin{array}{cccc|c}
1-\eta_i(1) & \eta_i(1) & 0 &  0 & 0\\
0 & 1-\eta_i(2) & \eta_i(2) & 0 & 0\\
0 & 0 & 1-\eta_i(3) & \eta_i(3)  & 0\\
0 & 0 & 0 & 1-\eta_i(4) & \eta_i(4)  \\
\hline
0 & 0 &0 & 0  & 1
\end{array}
\right)
\end{equation*}
where $\eta_i(k)=\mathbb{P}(R_i=k+1 | R_{i-1}=k)$ is a prior distribution. In order to obtain a valid segmentation of $n$ points into $K$ segments, one must add the constraint that $\{R_n=K\}$, this is why the model can be seen as a constrained HMM. A very natural choice for the prior distribution is to use $\eta_i(k) = \text{constant} \in [0,1]$ which leads to a uniform prior distribution over the space of segmentations. But more sophisticated prior might be use: priors forbidding change-points at certain locations (this might for example be useful for dealing with ties in data ordering), priors incorporating knowledge on most likely breakpoint locations, or even using posterior segmentation distribution from a previous study as a prior. 

For any given parameter $\boldsymbol{\theta}$, we then introduce the following forward and backward quantities: $F_i(k;\boldsymbol{\theta})=\mathbb{P}(\text{data}_{1:i},R_i=k ;\boldsymbol{\theta})$ and $B_i(k;\boldsymbol{\theta})=\mathbb{P}(\text{data}_{(i+1):n}, $ $R_n=K|R_i=k ;\boldsymbol{\theta})$ for all $i\in\{1,\ldots,n\}$ and $k \in \{1,\ldots,K\}$. These quantities can be computed recursively using the following recursions:
\begin{equation}\label{eq:fw}
F_i(k;\boldsymbol{\theta})=F_{i-1}(k-1;\boldsymbol{\theta})\eta_i(k-1)e_{i}(k;\boldsymbol{\theta})+F_{i-1}(k;\boldsymbol{\theta})(1-\eta_i(k))e_{i}(k;\boldsymbol{\theta}) 
\end{equation}
\begin{equation}\label{eq:bk}
B_{i-1}(k;\boldsymbol{\theta})=(1-\eta_i(k))e_{i}(k;\boldsymbol{\theta})B_i(k;\boldsymbol{\theta})+\eta_i(k)e_{i}(k+1;\boldsymbol{\theta})B_i(k+1;\boldsymbol{\theta})
\end{equation}
and we can derive from them posterior distributions of interest:
\begin{equation}\label{eq:posts}
\mathbb{P}(R_{i}=k | \text{data} ; \boldsymbol{\theta}) = w_i(k; \boldsymbol{\theta}) \propto F_i(k;\boldsymbol{\theta})B_i(k;\boldsymbol{\theta})
\end{equation}
\begin{equation}\label{eq:postbp}
\mathbb{P}(\text{BP}_k=i | \text{data} ;\boldsymbol{\theta}) \propto F_{i}(k;\boldsymbol{\theta})\eta_{i+1}(k)e_{i+1}(k+1;\boldsymbol{\theta})B_{i+1}(k+1;\boldsymbol{\theta})
\end{equation}
where $\{\text{BP}_k=i\}=\{R_{i}=k,R_{i+1}=k+1\}$. It is hence clear that Equation~(\ref{eq:posts}) allows to compute the marginal weights used in the EM algorithm (Section~\ref{EMalgo}) while Equation~(\ref{eq:postbp}) gives the marginal distribution of the $k^\text{th}$ breakpoint. Note that the full posterior segmentation distribution can be proved to be an heterogeneous Markov chain which transition can be derived immediatelty from Equations~(\ref{eq:posts}) and (\ref{eq:postbp}) \citep[see][for more details]{Nuel}.

Let us finally point out that the likelihood can also be derived from the forward-backward quantities and for any $i \in \{1,\ldots,n\}$ as:
\begin{equation}\label{eq:lik}
\mathbb{P}(\text{data} | \boldsymbol \theta) = \frac{\sum_{R_{1:n}} \mathbb{P}(\text{data} ,  R_{1:n} , R_n=K | \boldsymbol \theta)}
{\sum_{R_{1:n}} \mathbb{P}(R_{1:n} , R_n=K | \boldsymbol \theta)}
=\frac{\sum_{k=1}^K F_{i}(k;\boldsymbol{\theta})B_i(k;\boldsymbol{\theta})}
{\sum_{k=1}^K F_{i}^0(k)B_i^0(k)}
\end{equation}
where $F^0$ and $B^0$ are obtained through recursions (\ref{eq:fw}) and (\ref{eq:bk}) by replacing all $e_i(k;\boldsymbol{\theta})$ by $1$:
\begin{equation*}
F_i^0(k)=F_{i-1}^0(k-1)\eta_i(k-1)+F_{i-1}^0(k)(1-\eta_i(k)) 
\end{equation*}
\begin{equation*}
B_{i-1}^0(k)=(1-\eta_i(k))B_i^0(k)+\eta_i(k)B_i^0(k+1).
\end{equation*}
These quantities depend only on $\eta$, $n$ and $K$, thus they do not need to be updated during the EM algorithm.

In the (important) particular case where there is a uniform prior on the segmentation, one can use the constant $\eta_i(k)=\eta$. Simple combinatorics hence lead to $\sum_k F_i^0(k)B_i^0(k)= (1-\eta)^{n-K} \eta^{K-1}  {n-1 \choose K-1}$. Recursion can even be performed much faster by replacing all $\eta$ and $1-\eta$ by $1$ in all recursions. In this case, the probability distribution is defined up to a normalisation factor which is simply the binomial coefficient ${n-1 \choose K-1}$.

\section{Log-likelihood maximization with known weights}\label{sec:maxim}

Suppose you have at hand some preliminary estimator $\boldsymbol \theta_\text{old}$. 
 In Section~\ref{sec:posterior}, we showed how to use this quantity to estimate the marginal posterior probability $w_i(k;\boldsymbol\theta_\text{old})$ of position $i$ to be in the $k^{\text th}$ segment given the data and under $\boldsymbol\theta_\text{old}$. 
 From the expression of the $e_i(k,\boldsymbol{\theta})$ derived in~\eqref{likelihood}, Equation~\eqref{EM} can be solved by maximizing a simple weighted log-likelihood. When the weights are all equal to $1$, statistical inference has already been studied, either in a fully parametric case if one assumes a parametric form for the baseline hazard rate \citep[see for instance][]{kalbfleisch_book} or in a semiparametric way if the baseline hazard rate is left unspecified which corresponds to the well known Cox model. In the latter case, a weighted log-likelihood has also been briefly studied in~\cite{therneau}, pages 161-168. But in both parametric and semiparametric cases, our weighted log-likelihood estimation procedure is very similar to the standard estimation techniques used in the absence of weights. 


%
In the next section, we present the implementation of our estimator for different choices for the baseline hazard rate in a Cox model. We propose to use either a parametric baseline among the exponential, the Weibull and the piecewise constant hazard or to use a nonparametric baseline, that is to let the baseline hazard unspecified. The expression of the different families for the baseline hazard are all recalled in the Supporting Material. The piecewise constant hazard model is very useful when one does not know the shape of the baseline hazard a priori. However it requires to choose a pre specified number of cutpoints. 
The nonparametric case is the most flexible model since it does not require any particular form for the baseline hazard. In the classical Cox model, the Cox's partial likelihood provides efficient estimation of the regression parameters and estimation of the cumulative baseline is performed through the Breslow estimator \citep[see][]{Breslow1972}. However, in our context, classical estimation methods will not lead to consistent estimators due to numerical instabilities. In order to consistently estimate the model parameter and the posterior segment distribution with a nonparametric baseline, a smooth estimator of the baseline is required. This is introduced in Section~\ref{smoothing}.
 Choice of the number of cutpoints in the piecewise constant hazard model and choice of the bandwidth in the nonparametric case are discussed in Section~\ref{sec:discussion}.
%

\section{Practical implementation}\label{sec:global}

\subsection{Parametric baseline hazards}

The parametric case is straightforward: the final estimators are obtained by alternating computation of the estimates through Equation~\eqref{EM} and computation of the weights through the posterior segment distribution calculated in Section~\ref{sec:posterior}. 

The algorithm of our estimation procedure is as follows. First suppose you have at your disposal an initial weight function $w_i(k;\boldsymbol\theta_\text{old})$.

\begin{enumerate}
\item [Step 1.] Compute $\widehat{\boldsymbol{\theta}}=\argmax_{\boldsymbol \theta } \,Q(\boldsymbol{\theta} | \boldsymbol{\theta}_\text{old})$ from Equation~\eqref{EM}. In the exponential or Weibull models, this can be done via the \textbf{survreg} function in R (see Section 2 of the Supporting Material) and in the piecewise constant hazard model, this can be done via the \textbf{glm} function in R (see Section 3 of the Supporting Material)
\item [Step 2.] Compute the new weights $w_i(k;\widehat{\boldsymbol\theta})$ using Equation~\eqref{eq:posts} in Section~\ref{sec:posterior}.
\item [Step 3.] Let $\boldsymbol\theta_\text{old}=\widehat{\boldsymbol\theta}$ and return to Step 1.
\end{enumerate}

\subsection{Nonparametric baseline hazard}\label{smoothing}

The nonparametric case requires one supplementary step. After the first step, smoothed versions of the baseline hazard and cumulative baseline hazard estimators need to be derived. The weighted log-likelihood and the weights are then computed using these smoothed estimators. We propose in this work to use kernel type estimators but our method could be extended to any type of smoothing estimators such as wavelets, splines, k-nearest neighbor estimators, projection estimators etc. 

Let $\mathcal K$ be a kernel such that $\int \mathcal K(u)du=1$, $\int u \mathcal K(u)du=0$, $\int u^2 \mathcal K(u)du<\infty$ and $\int \mathcal K^2(u)du<\infty$. Let $h$ be a bandwidth satisfying $h\to 0$ and $nh\to \infty$ as $n$ tends to infinity. We note $\tilde \Lambda_k$ the estimator of $\Lambda_k$ obtained from the weighted Cox partial likelihood (see the Supporting Material for an explicit expression of this estimator) and we introduce smoothed estimators of $\lambda_k$:
\begin{align}\label{smooth}
\hat\lambda_k(t)=\frac{1}{h}\sum_{i=1}^n\int \mathcal K\left(\frac{u-t}{h}\right)d\tilde\Lambda_k(u)\text{ and } \hat\Lambda_k(t)= \int_0^t \hat\lambda_k(s)ds.
\end{align}
Let $\widehat{\boldsymbol\theta}=(\hat\Lambda_1,\ldots,\hat\Lambda_K, \widehat{\boldsymbol\beta}_1,\ldots,\widehat{\boldsymbol\beta}_K)$. This new estimator is now used to estimate $e_i(k;\boldsymbol{\theta})$ and then to obtain estimators of the weights. From Equation~\eqref{likelihood} we have:
\begin{align}\label{likelihood2}
\log \left(e_i(k;\widehat {\boldsymbol{\theta}})\right)=\Delta_i\left(\log\big(\hat\lambda_k(T_i)\big)+\X_i\widehat{\boldsymbol\beta}_k\right)- \hat\Lambda_k(T_i)\exp(\X_i\widehat{\boldsymbol\beta}_k). 
\end{align}
Note that the weighted likelihood $Q(\widehat{\boldsymbol\theta}|\boldsymbol\theta_{\text {old}})$ obtained from these $e_i(k;\widehat {\boldsymbol{\theta}})$ does not reduce to a partial likelihood due to the use of smoothed hazard and cumulative hazard estimators. However this is not an important matter since our algorithm does not require the maximization of this likelihood: Equation~\eqref{likelihood2} is only needed for the computation of the new weights from Equation~\eqref{eq:posts} in Section~\ref{sec:posterior} while the optimization step only involves the Cox partial likelihood and is easily performed through the Newton-Raphson algorithm.

The final algorithm of our estimation procedure is as follows. First suppose you have at your disposal an initial weight function $w_i(k;\boldsymbol\theta_\text{old})$.
\begin{enumerate}
\item [Step 1.] Compute $\widetilde{\boldsymbol{\theta}}$ using the Newton-Raphson algorithm to maximize the weighted Cox partial likelihood (see the Supporting Material for details about the Newton-Raphson algorithm). This can be done via the \textbf{coxph} function in R with a weight option. 
\item [Step 2.] Smooth the $\tilde\lambda_k$ and $\tilde\Lambda_k$ using Equation~\eqref{smooth}. This gives $\widehat{\boldsymbol{\theta}}$.
\item [Step 3.] Compute $\log \left(e_i(k;\widehat {\boldsymbol{\theta}})\right)$ as in Equation~\eqref{likelihood2} and get the new weights $w_i(k;\widehat{\boldsymbol\theta})$ from Equation~\eqref{eq:posts} in Section~\ref{sec:posterior}. 
\item [Step 4.] Let $\boldsymbol\theta_\text{old}=\widehat{\boldsymbol\theta}$ and return to Step 1.
\end{enumerate}

\subsection{Choice of the parameters and stopping rule to find the correct model}\label{sec:discussion}

These algorithms need to be initialized by either choosing initial model parameters or by directly choosing initial weights $w$. We propose the following \emph{ad-hoc} method to initialize the weights for a sample of size $n$ and $K$ segments. 
First divide the sample in $K$ segments and for any individual $i$ in segment $k$, choose $w_i(k,\boldsymbol \theta_\text{old})=w$ with $w$ a high number between $0$ and $1$ (for instance, take $w=0.7$). For any individual $j$ that is not in segment $k$, choose $w_j(k,\boldsymbol \theta_\text{old})=1-w$. 

In all models, the Newton-Raphson algorithm is initialized by taking the null vector for $\widehat{\boldsymbol\beta}_k^{(0)}$. Step $2$ in the parametric models and step 3 in the Cox model are performed using the R package \texttt{postCP} developed by~\cite{Nuel}.

The exponential and Weibull baseline hazard models only require the initialization of either the model parameters or the weights. On the opposite, the piecewise constant baseline hazard model and the nonparametric baseline model require an extra parameter to be chosen. In both models, the estimation procedure is not very sensitive to the choice of this parameter, especially in terms of breakpoints detection. In particular, the number of cut points in the piecewise constant hazard is set by default to $3$ and as shown in the simulation section, this leads to very performant breakpoints selection. Increasing the number of cut points does usually not make the breakpoints detection more accurate. These $3$ breakpoints can be chosen for instance from the data as the quantiles of the event times of order $0.25$, $0.5$ and $0.75$ respectively. The same phenomena happens for the choice of the bandwidth in the nonparametric model: detecting the correct number of breakpoints is not much affected by the choice of the bandwidth. However, it might still be of interest to find an optimal bandwidth if one wants to give a precise estimation of the baseline hazard. This problem is classical for density estimation and has been studied for nonparametric estimation of baseline hazards by~\cite{ABGK}. Equations (4.2.25) and (4.2.26) of their book suggest that a bandwidth of order $n^{-1/5}$ would give the best compromise between bias and variance trade-off in the estimation of the baseline hazard. In particular asymptotic normality of order $(nh)^{1/2}$ would be achieved with such a bandwidth as expressed by their theorem IV.2.4. More discussions about how to choose the bandwidth from the data can be found in~\cite{ABGK}, see in particular their Examples IV.2.3, IV.2.4 and IV.2.5. Since the interest in the choice of the bandwidth is limited in our context we will not pursue this discussion here but as a rule of thumb we recommend the user to choose $h=n^{-1/5}$ in real data situations.


One other important issue is to find the correct number of breakpoints in the dataset. A simple solution consists to start with a model with one breakpoint and increment the number of breakpoints one by one. As presented in the real data analysis for example (see Section~\ref{data}) a visual inspection of the plots of the maximum a posteriori of the breakpoints can help to find the right model. However, the conclusion from theses plots can be subjective and it is therefore important to propose a numerical indicator that helps discriminating between different models.  We propose the following BIC criterion designed to make a tradeoff between information provided by the data on a model and the complexity of the model:
\begin{equation*}
\text{BIC}(d)=-2 \log \mathbb{P}(\text{data} | \boldsymbol{\hat\theta})+d\log(n)
\end{equation*}
where the likelihood $\mathbb{P}(\text{data} | \boldsymbol{\hat\theta})$ can be computed using Equation~(\ref{eq:lik}), and $d$ corresponds to the dimension of the model. The value of $d$ is different for every model, it corresponds to the total number of parameters that need to be estimated. For the exponential baseline, $d=(p+1)K$, for the Weibull baseline, $d=(p+2)K$ and for the piecewise constant hazard baseline, $d=(p+L)K$. No such indicator can be derived for the nonparametric baseline hazard since in that case the number of parameters to be estimated equals infinity. This BIC criterion is used in Section~\ref{data} for the exponential baseline to discriminate between different models and find the correct number of breakpoints.

%

\section{Simulated data}\label{sec:simu}

In this section we evaluate the performance of our estimation technique through numerical experiments. We consider a Cox model as defined by Equation~\eqref{model}, with $K=3$ segments and a binary covariate $\boldsymbol X$ distributed as a Bernoulli variable with parameter equal to $0.5$. We consider different scenarios corresponding to different baseline hazards and different regression parameters:

\begin{enumerate}
\item [Scenario 1.] Exponential baselines, $\lambda_1(t)=1$, $\lambda_2(t)=0.5$, $\lambda_3(t)=0.7$ and $\beta_1=1.5$, $\beta_2=-0.5$, $\beta_3=-0.5$.
\item [Scenario 2.] Weibull baselines, $\lambda_1(t)=5t^{4}$, $\lambda_2(t)=2t$, $\lambda_3(t)=2t$ and $\beta_1=1.5$, $\beta_2=-1$, $\beta_3=-5$.
\item [Scenario 3.] Piecewise constant baselines, 
\begin{align*}
\lambda_1(t) & =0.8\,I(0<t\leq 1)+1.2\,I(1<t\leq 3)+1.6\,I(3<t),\\
\lambda_2(t)&= 1.2\,I(0<t\leq 4)+1.6\,I(4<t\leq 6)+2\,I(6<t),\\
\lambda_3(t)&=1.6\,I(0<t\leq 5)+2\,I(5<t\leq 7)+2.4\,I(7<t), 
\end{align*}
and $\beta_1=1.5$, $\beta_2=-0.5$, $\beta_3=-1.5$.
\item [Scenario 4.] Gompertz baselines, $\lambda_1(t)=e^{5t}$, $\lambda_2(t)=e^{2t}$, $\lambda_3(t)=e^{2t}$ and $\beta_1=1.5$, $\beta_2=-0.5$, $\beta_3=-1.5$.
\end{enumerate}

\begin{figure}[!p]
\begin{tabular}{cc}
\includegraphics[width=0.40\textwidth]{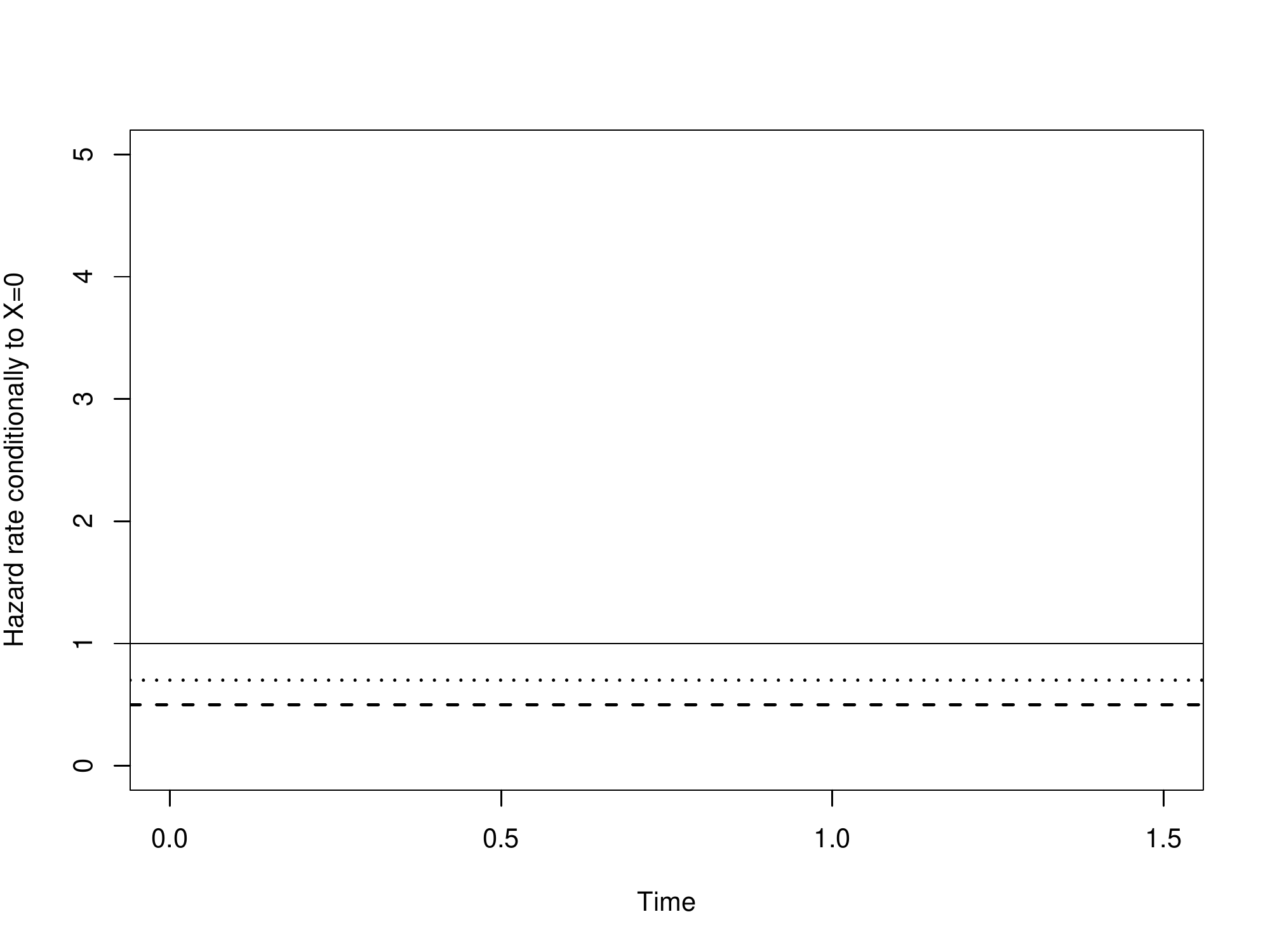}&\includegraphics[width=0.40\textwidth]{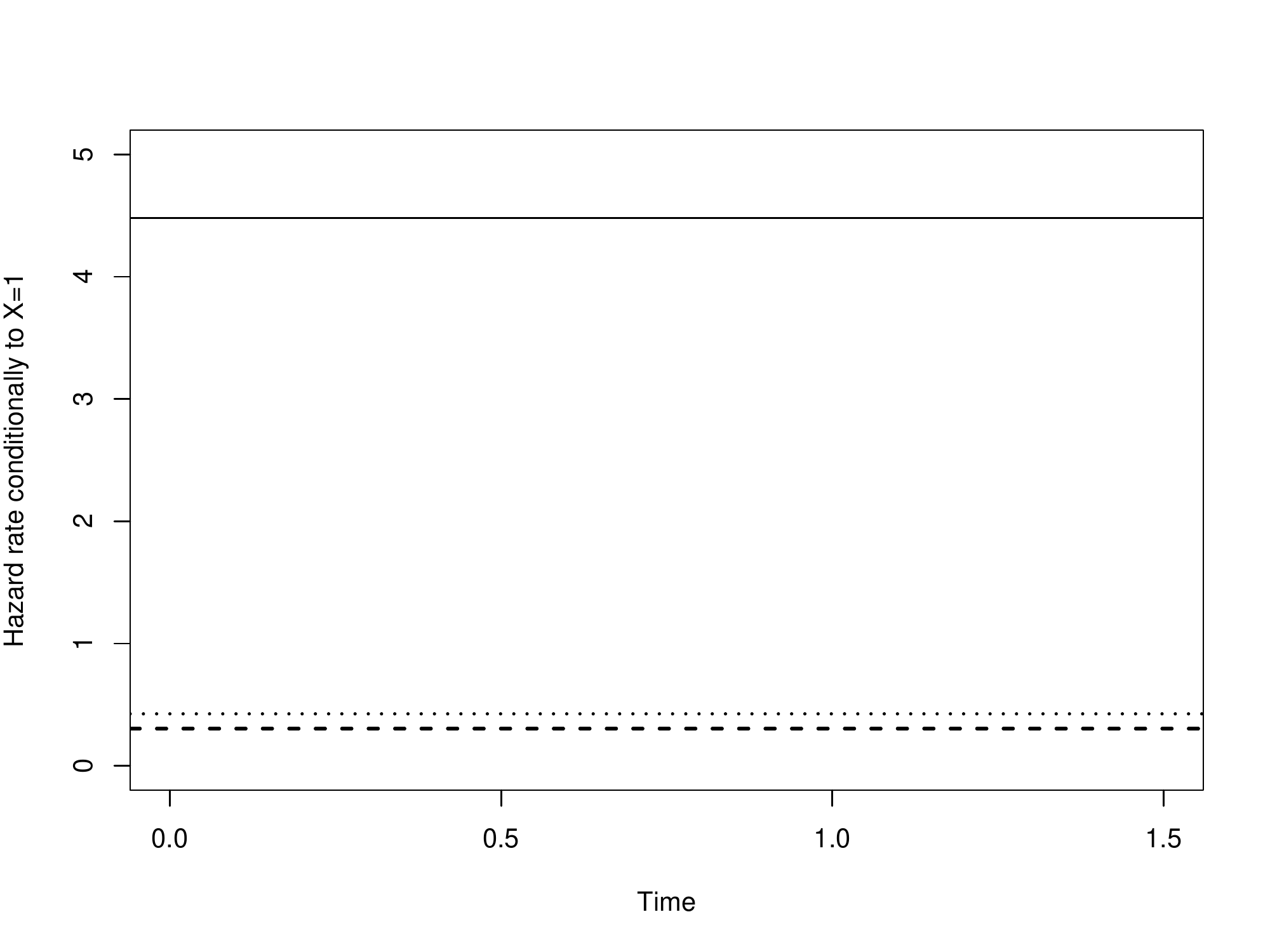}\\
\includegraphics[width=0.40\textwidth]{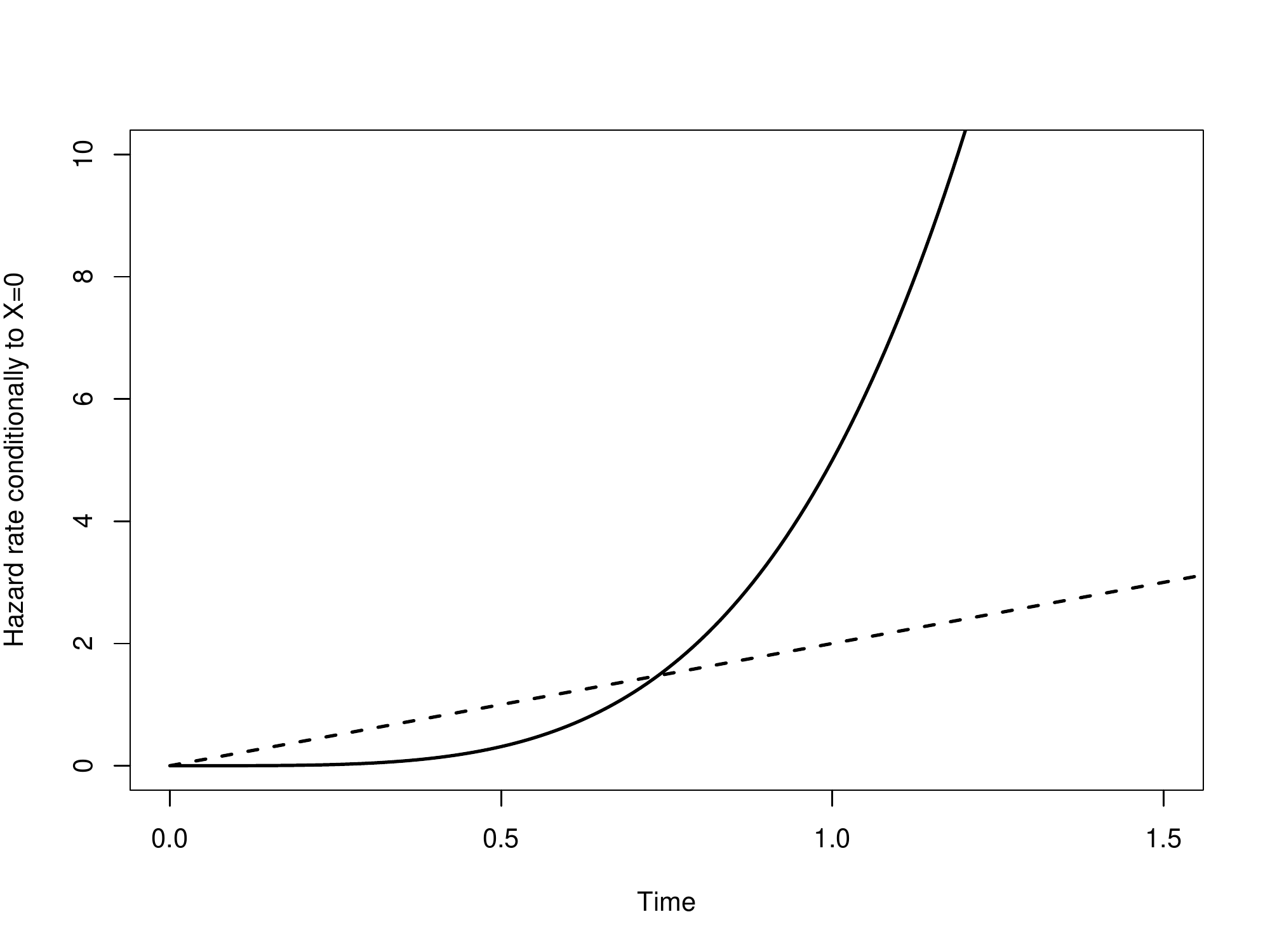}&\includegraphics[width=0.40\textwidth]{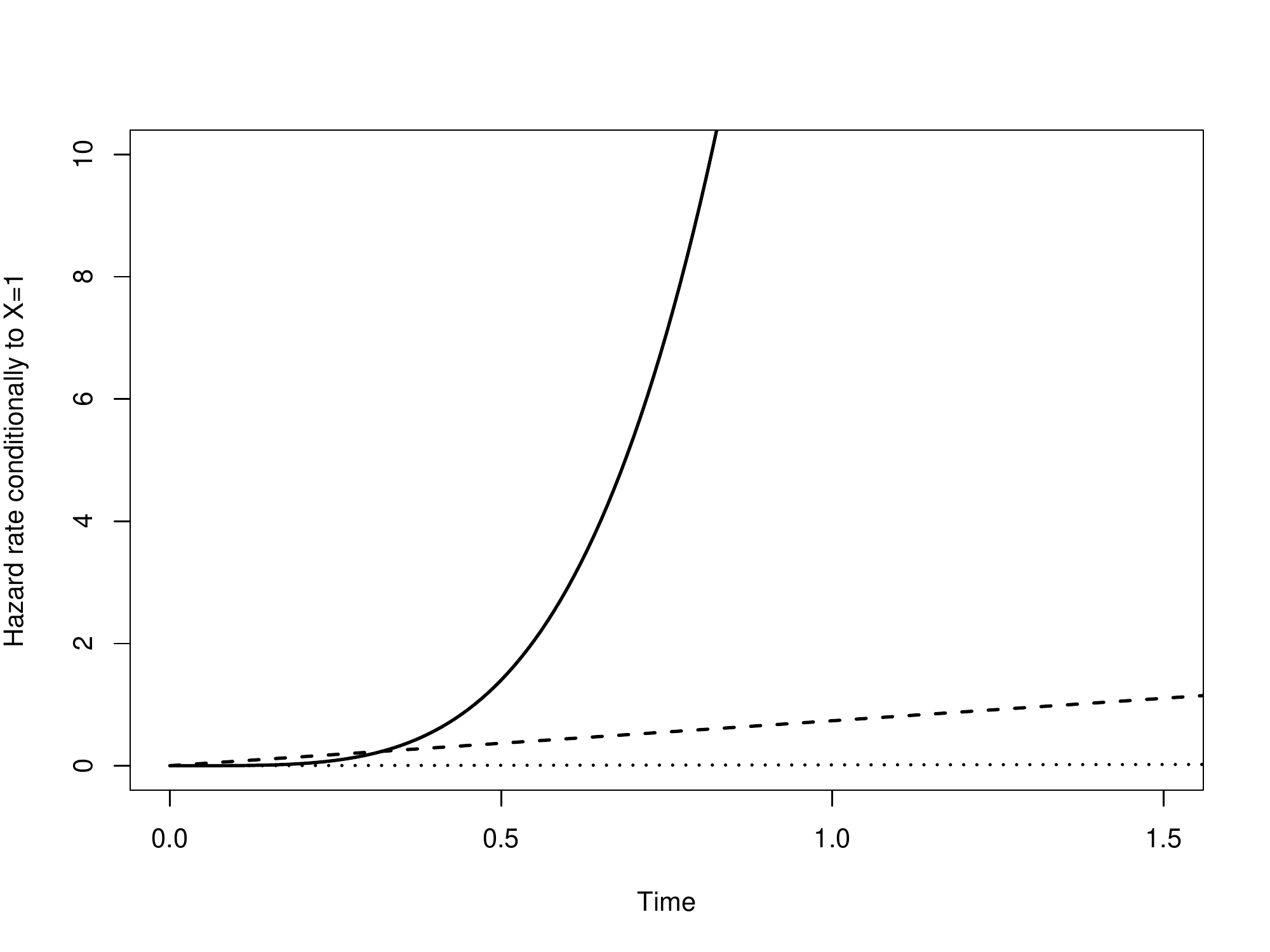}\\
\includegraphics[width=0.40\textwidth]{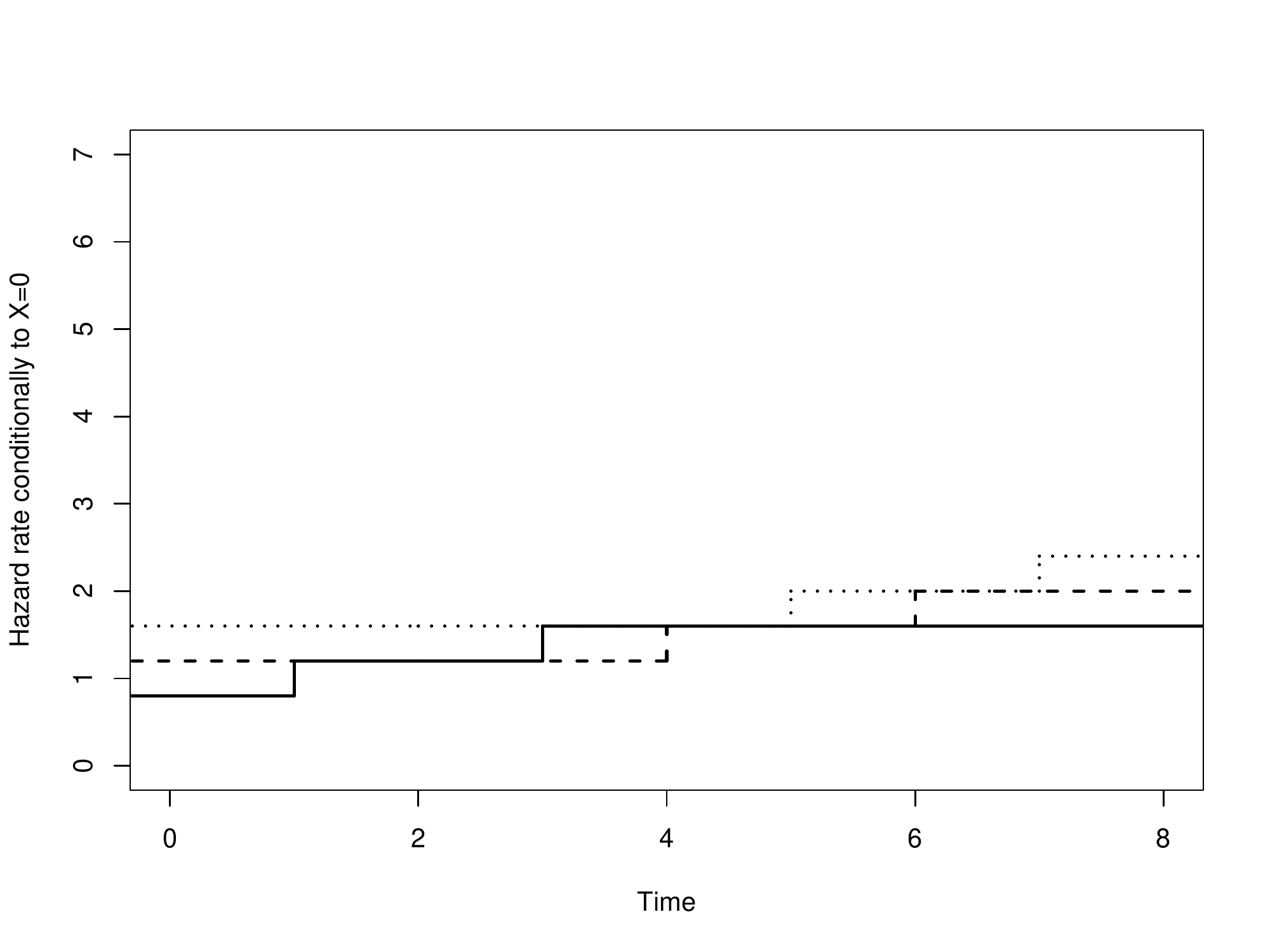}&\includegraphics[width=0.40\textwidth]{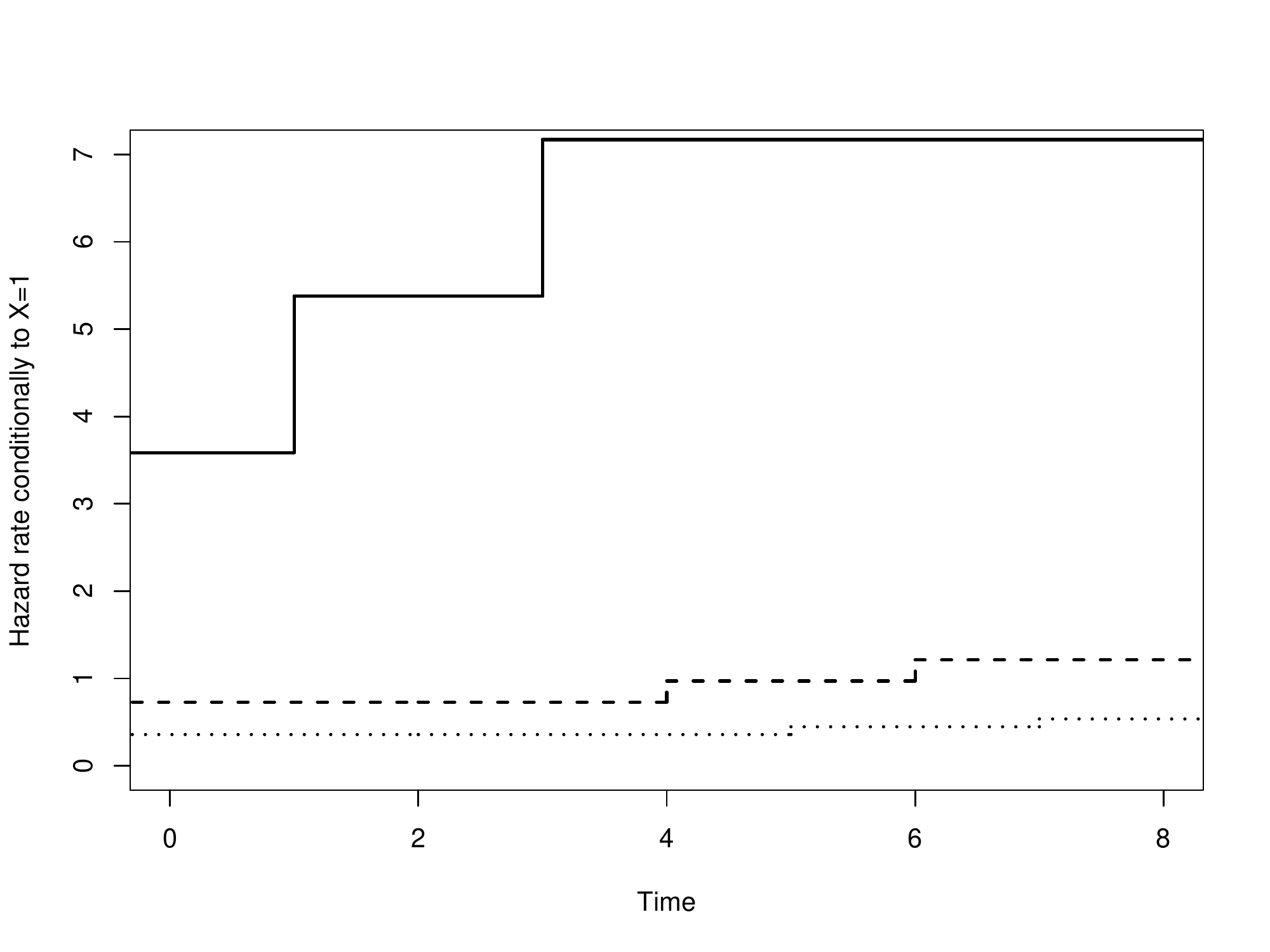}\\
\includegraphics[width=0.40\textwidth]{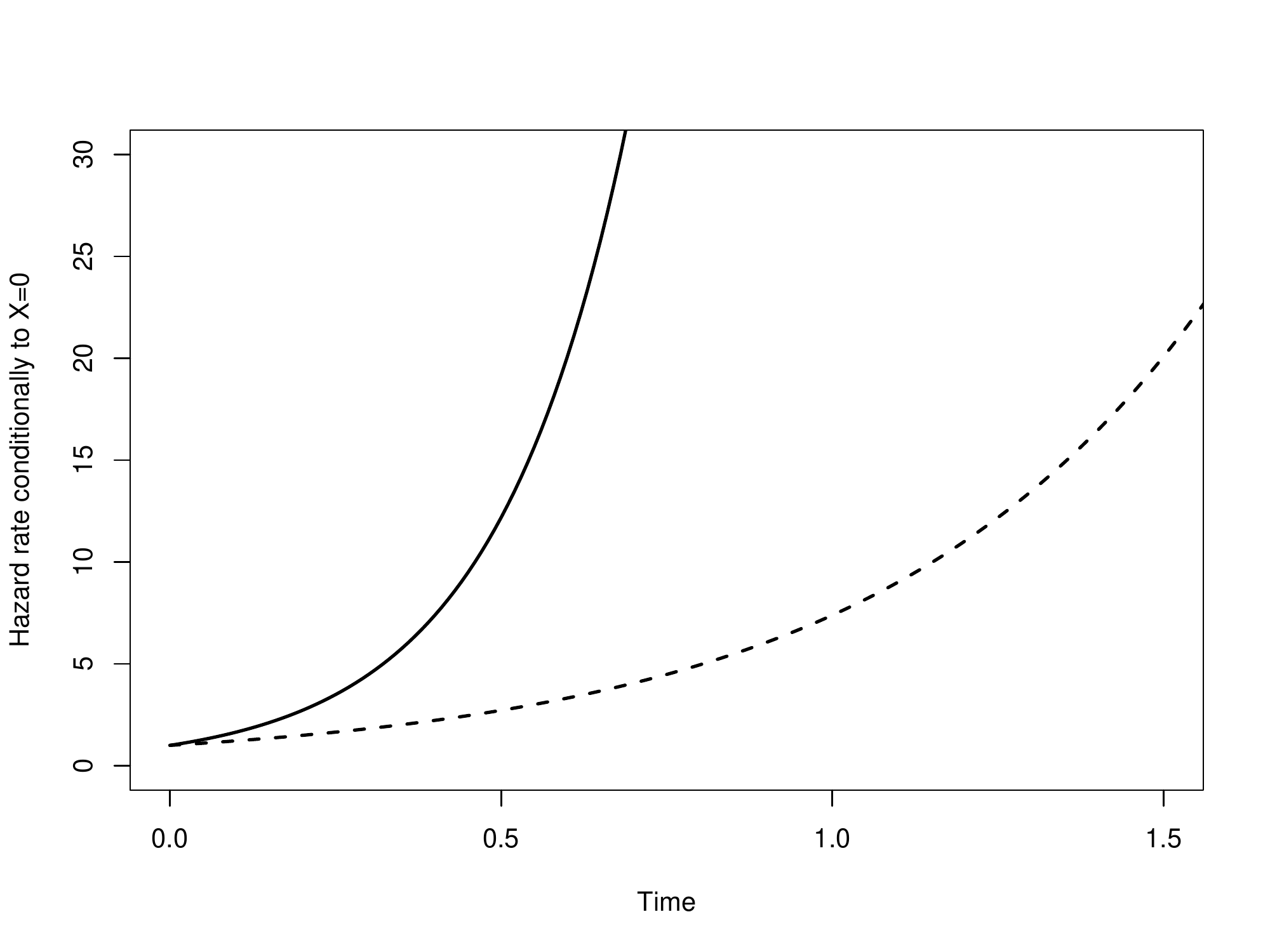}&\includegraphics[width=0.40\textwidth]{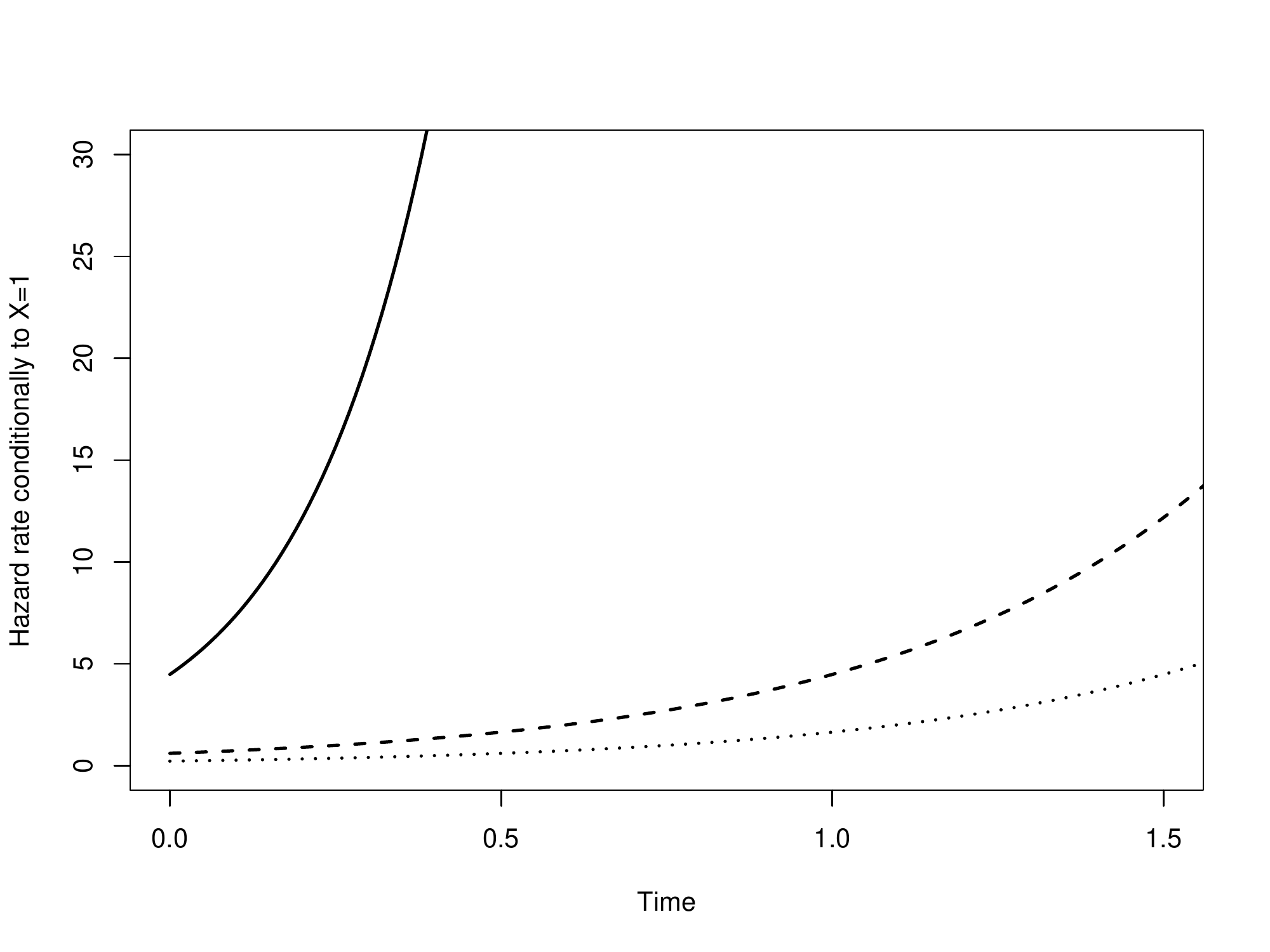}\\
\end{tabular}
\caption{Conditional hazard rates in simulated data for Scenarios 1 to 4 from top to bottom. Solid line: hazard in segment 1. Dash line: hazard rate in Segment 2. Dot line: hazard rate in Segment 3.}
\label{hazards}
\end{figure}

In all four scenarios, the sample size $n$ equals $3\,000$, and the data were simulated such that $R_{1}=\cdots=R_{1000}=1$, $R_{1001}=\cdots=R_{2000}=2$ and $R_{2001}=\cdots=R_{3000}=3$. Each scenario was calibrated such that the change in the hazard distribution between Segments $1$ and $2$ was more important than the difference in the hazard distribution between Segments $2$ and $3$. This is illustrated by Figure~\ref{hazards} which provides the plots of the conditional hazard rates in each scenario. The censoring variable was chosen as a uniform distribution such that approximately $50\%$ of the observations were censored in each scenario. Exact parameter values of the censoring distribution can be found in the Supporting Material. 
For the piecewise constant hazard model estimator, as recommended in Section~\ref{sec:discussion}, the cuts positions were chosen from the empirical quantiles of order $0.25$, $0.5$ and $0.75$ of the data. This lead us to take the approximate values $0.2, 0.5$ and $1.1$ for Scenario 1, $0.4, 0.7, 1$ for Scenario 2, $0.15$, $0.35$ and $0.5$ for Scenario 3 and $0.1$, $0.2$ and $0.4$ for Scenario 4. For the nonparametric baseline hazard model estimator, as recommended in Section~\ref{sec:discussion}, the bandwidth was chosen equals to $3000^{-1/5}\approx0.2$ in all scenarios. Finally we ran $1\,000$ replications of each of these scenarios and the results were reported in Table~\ref{Table1}. Following formula~\eqref{eq:postbp} the maximum a posteriori of a breakpoint was computed on each Monte-Carlo sample and the mean location and mean value of that maximum were reported in Table~\ref{Table1}. Empirical confidence intervals were also computed for this maximum a posteriori of breakpoint.

 In all scenarios, detection of the first breakpoint is usually very accurate where in many cases the average breakpoint location is exactly equal to the true breakpoint location, $1\,000$. The second breakpoint is more difficult to detect as shown by wider confidence intervals even though the average breakpoint location is usually close to the true breakpoint location, $2\,000$. The average value of the marginal probability of breakpoint detections also illustrate the uncertainty about the second breakpoint location: the probability for the first breakpoint location is in all cases much higher than for the second breakpoint location.
 
The most problematic breakpoint to find corresponds to the breakpoint from segment $2$ to $3$ under Scenario 1 and as a matter of fact none of the proposed methods manage to provide an accurate $95\%$ confidence interval. In this scenario, for every estimation methods there was a probability of approximately $1$ over $1\,000$ that the algorithm fails to find the second breakpoints leading to an error in the program.

It is interesting to notice that on the overall the true hazard distribution of the data does not seem to play any role in the detection power of our estimation methods as long as the change in the hazard distribution in two segments is large enough. For instance, in Scenario 4, which involves a simulation setup that does not correspond to any of the parametric baseline distributions proposed in the different estimation methods, all estimators find very accurate breakpoint locations with very narrow confidence intervals. The estimation performance of the regression parameter does not seem to be much affected by the data simulation setup neither, since the Weibull, piecewise constant and nonparametric baseline estimators show little difference in their estimation performance from one scenario to another. 
One exception is the exponential baseline estimator which seems to behave poorly in Scenarios 2 and 4 when looking at the regression parameter estimates and the confidence intervals for the second breakpoint compared to the other estimators.  

Globally, all estimators are performant both in breakpoint detections and parameters estimation as long as the change in the hazard distribution is big enough from one segment to another. In that case, the nonparametric baseline estimator seems to give the biggest value of the probability of the breakpoint distribution. 
When only a slight change occurs between the hazard distribution of two segments, all the proposed methods are less precise and the exponential baseline estimator seems to be the less performant of all baseline estimators.




More simulation studies which are not reported here have been carried out. When the change in distribution between two segments increases, the probability of the marginal breakpoint distribution increases accordingly and can be almost equal to $1$ in some situations. For instance the marginal probabilities of the breakpoints found in Section~\ref{data} seem to indicate a much drastic change in the survival distributions than the simulation setting presented here. Finally scenarios with a mixture of different parametric survival distributions in each segment have also been investigated. These simulations lead to similar behaviour of our estimators and are therefore omitted.

\section{Robustness study for the BIC criterion}\label{sec:robust}

\begin{figure}[!p]
\includegraphics[height=3.8in,width=6in]{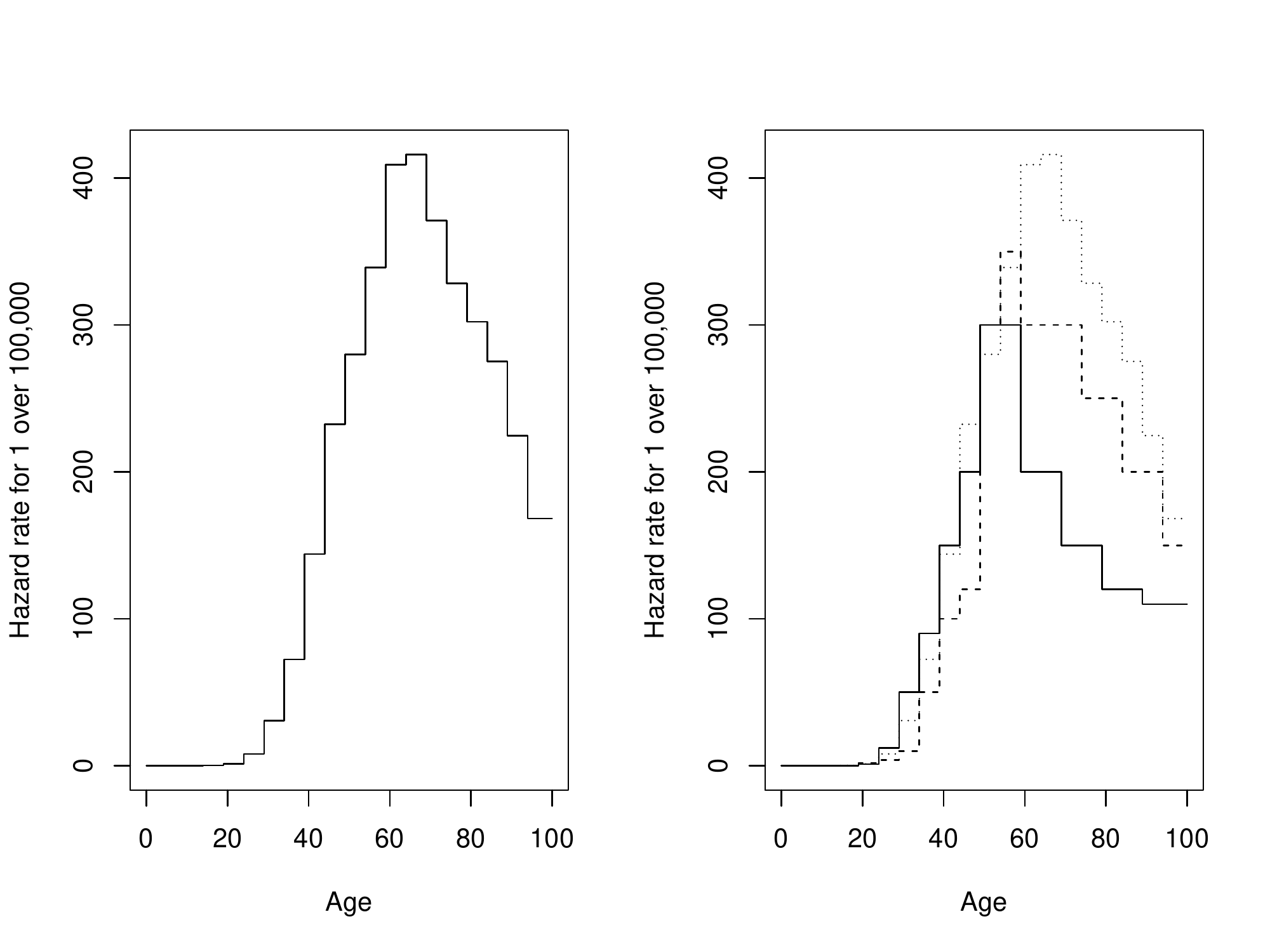}\\
\caption{Left panel: simulated hazard rates for the null case (no breakpoints) based on the French national breast cancer incidence data. Right panel: simulated hazard rates for the two breakpoints model. Solid line: individuals $1$ to $15,000$. Dash line: individuals $15,001$ to $25,000$. Dot line: individuals $25,001$ to $35,000$.}
\label{cancer}
\end{figure}

In this section we evaluate how performant the BIC criterion is to choose the correct number of breakpoints. We propose two scenarios: a null case where there is no breakpoint in the population and an other case where there are two breakpoints. In order to make the comparison more realistic we simulated the data under the null case by mimicking the French national breast cancer incidence \citep{invs2015}. The two breakpoints simulation was obtained by adding a small noise to this null case. For the null case we simulated a sample of size $15,000$ and for the second scenario we simulated a sample of size $35,000$ with breakpoints at positions $15,000$ and $25,000$. The simulation was easily performed from a piecewise constant hazard model by choosing cuts of $5$ years length, starting from age $15$ until age $95$. The hazard curves are represented by Figure~\ref{cancer}.

We also studied the AIC criterion whose definition is similar to the BIC criterion but $\log(n)$ is replaced by the constant $2$. We computed these two indicators for a number of breakpoints ranging from $K=1$ to $K=6$ and computed the proportion of selected models for $1,000$ replications in each scenario using either the exponential baseline estimator or the piecewise constant hazard baseline estimator. The cuts for the piecewise constant hazard baseline estimator where found by taking the quantiles of order $0.25$, $0.5$ and $0.75$ as described in Section~\ref{sec:discussion}. Table~\ref{null} presents the results for the null case and the two breakpoints model.

Under the null case it is interesting to note that the BIC criterion will never find a breakpoint when there are none in the population. On the contrary, using the piecewise constant hazard baseline estimator, the AIC criterion will have approximately $8\%$ of chances to choose a breakpoint model when there are none. Also, in the two breakpoints scenario the BIC criterion gives clearly a much accurate prevision of the number of breakpoints compared to the AIC criterion. For the BIC criterion, the exponential baseline estimator seems to outperform the piecewise constant hazard baseline estimator since this estimator gives $98.7\%$ chances of finding the correct model as opposed to $92.9\%$ for the piecewise constant hazard baseline estimator.


\section{Survival analysis of diabetic patients at the Steno memorial hospital}\label{data}

In this section we illustrate our method on a dataset on survival of diabetics patients at the Steno memorial hospital. The data are described in great details in Example I.3.11 in~\cite{ABGK} and were originally studied through a illness-death model where the illness state corresponded to the diabetic nephropathy status of the patients. Here, we will only focus our interest on the survival of the patients, that is the variable of interest is the time from diagnosis of diabetes of a patient until death. The data were collected between $1933$ and $1981$ and patients were included in the study if the diagnosis of diabetes mellitus was established before age 31 years and between 1933 and 1972. A total of $2\,709$ patients were followed from the first contact with the hospital until death, emigration or the 31st of December 1984. On these $2\,709$ patients $707\, (26\%)$ deaths were observed and the other $2\,002\, (74\%)$ patients were considered right censored. Since most of the patients did not contact the hospital directly after the diagnosis of diabetes, patients in this dataset are also left truncated. This needs to be taken into account because it means that individuals have a delayed entry into the study and will be observed only if they did not die before attending the Steno hospital. Without appropriate methods to deal with left truncation our estimation techniques will tend to overestimate the survival of diabetics patients.  
Gender (coded as $0$ for women and $1$ for men) and the year of birth were recorded for every patients. The dataset is composed of approximately $56\%$ of male and $44\%$ of female. The years of birth range from $1903$ to $1971$ and the calendar year of onset of diabetes range from $1933$ to $1972$. Our aim was to determine if there was any change in the hazard distribution according to the calendar year of onset of diabetes when adjusting by gender. The marginal survival curves and parameter estimates in a Cox model with exponential baseline hazard were also computed. Finally a bootstrap procedure was implemented to provide valid confidence intervals that take into account all the variability in the estimation procedure coming from the location of the breakpoints, which is unknown and from the parameter estimates.

To accommodate our method for left truncation the individual at risk process $Y_i(t)$ needs to be replaced by $Y_i(t)=I(L_i\leq t\leq T_i)$ where $L_i$ represents the left truncation variable for individual $i$. This will affect the value of the emission probability $e_i(k;\boldsymbol{\theta})$ (see Equation~\eqref{likelihood}) which in turn will affect the value of the a posteriori segment distribution $w_i(k; \boldsymbol{\theta})$ and the value of the weighted log likelihood $Q(\boldsymbol{\theta} | \boldsymbol{\theta}_\text{old})$. The parameters are estimated by maximizing the log likelihood in Equation~\eqref{EM} as before. For example, in the exponential model, the logarithm of the emission probability is equal to:
\begin{align*}
\log e_i(k;\boldsymbol{\theta}) =  \Delta_i\left(-\log(\lambda_k)+{\boldsymbol X}_i{\boldsymbol\beta}_k\right)- \left(\frac{T_i-L_i}{\lambda_k}\right)\exp\left({\boldsymbol X}_i{\boldsymbol\beta}_k\right).
\end{align*}

Since only the year of diabetes onset (and not the exact date) it means 
that a breakpoint can only occur when changing from one year to another. To take this into account we first ordered all individuals with respect to their calendar year of diabetes onset and the computation of the posterior distribution was constrained through the priors $\eta_{i}(k)$, defined in Section~\ref{sec:posterior}, such that $\eta_{i}(k)=0$ for any $k$ if individuals $i$ and $i+1$ were diagnosed diabetics the same year. Other priors were set to $0.5$. Since $0$ is an absorbing state this ensured us to have change-points only for a new diabetes onset year.

Based on the results of Section~\ref{sec:robust} we decided to use the exponential baseline model to perform the estimation of the model parameters and to use the BIC criterion to find the correct number of breakpoints for a number of possible breakpoints ranging from zero to four. 

\begin{figure}[!p]
\begin{tabular}{cc}
\includegraphics[width=0.5\textwidth]{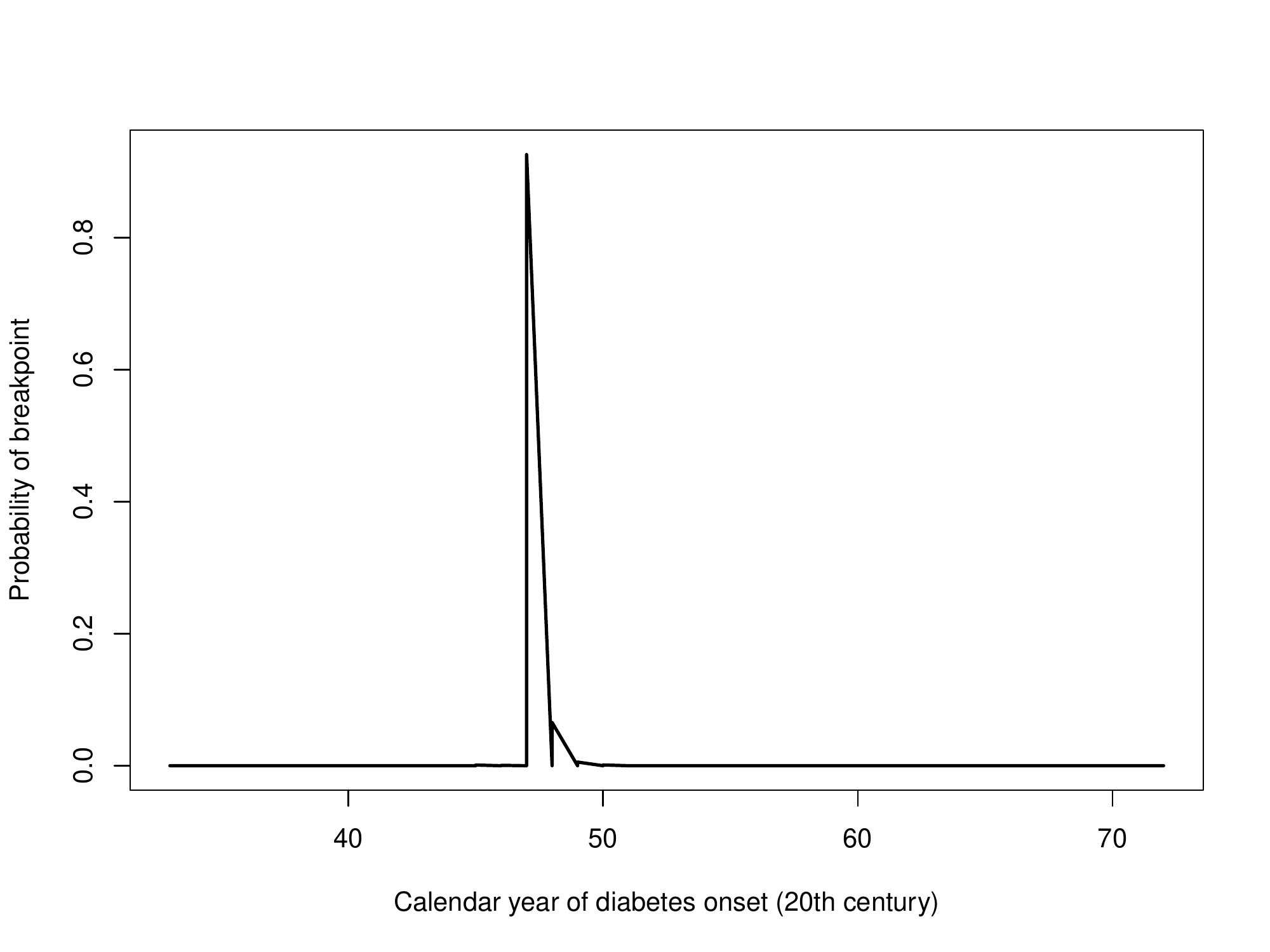}&\includegraphics[width=0.5\textwidth]{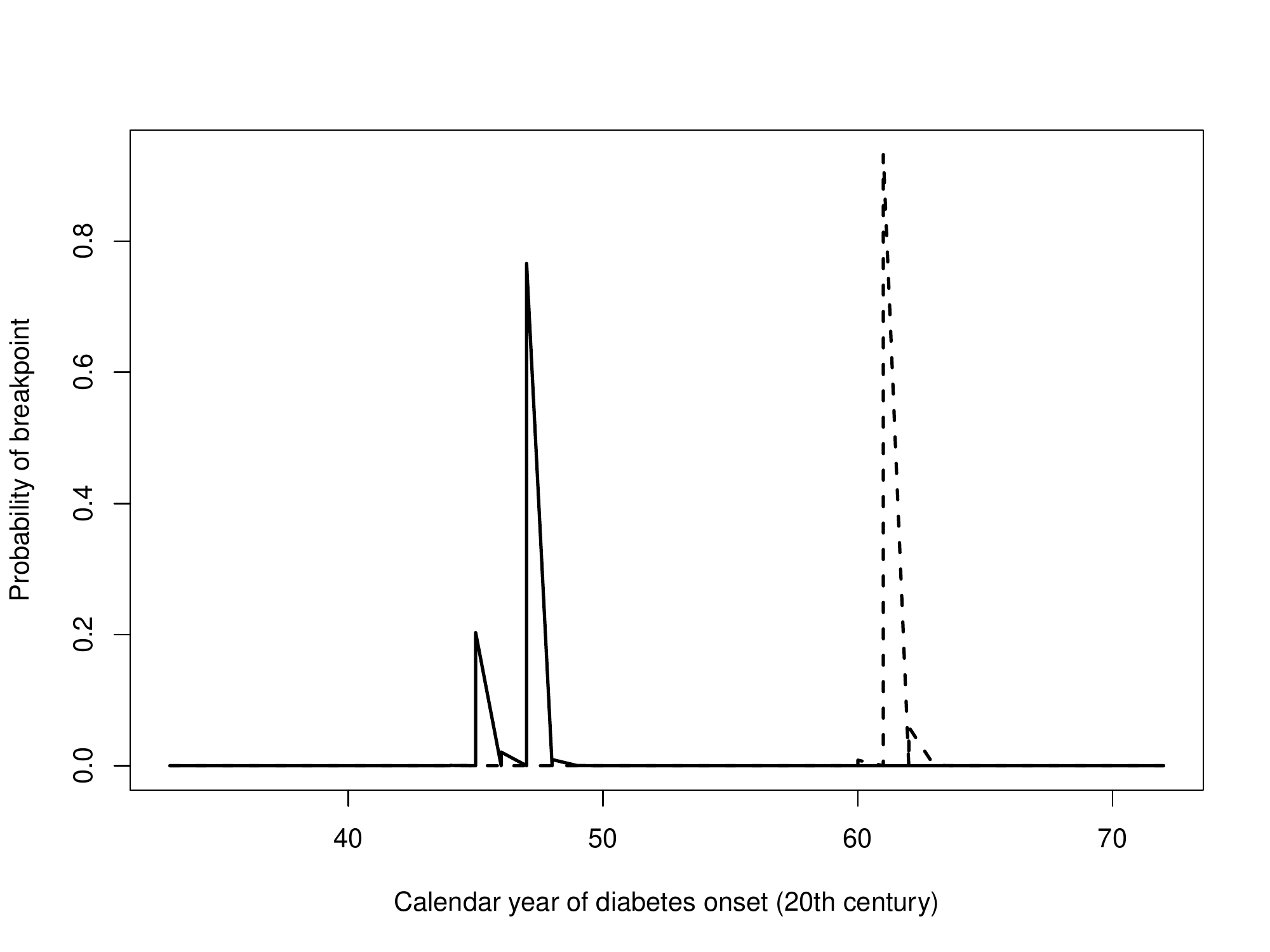}\\
\includegraphics[width=0.5\textwidth]{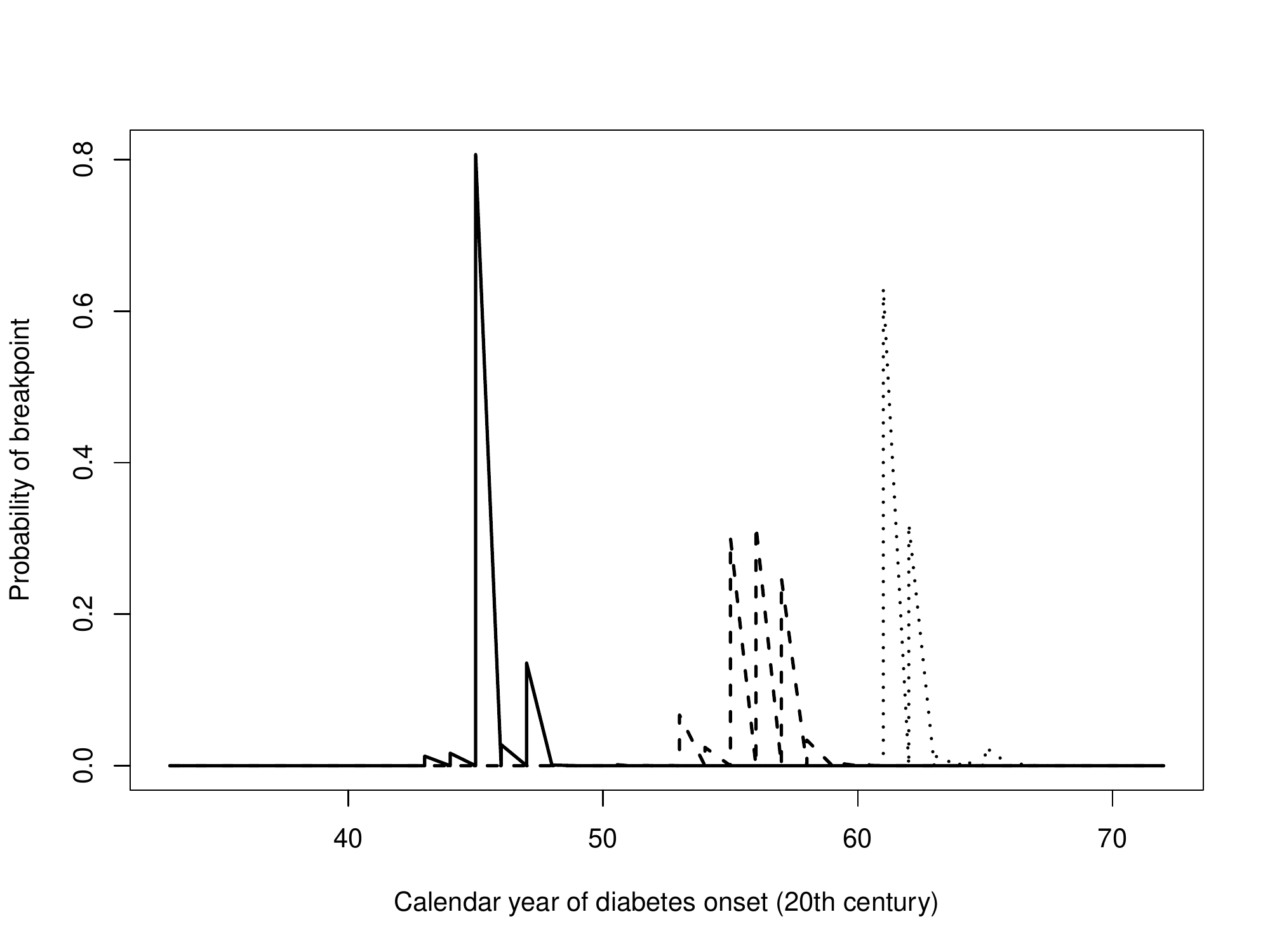}&\includegraphics[width=0.5\textwidth]{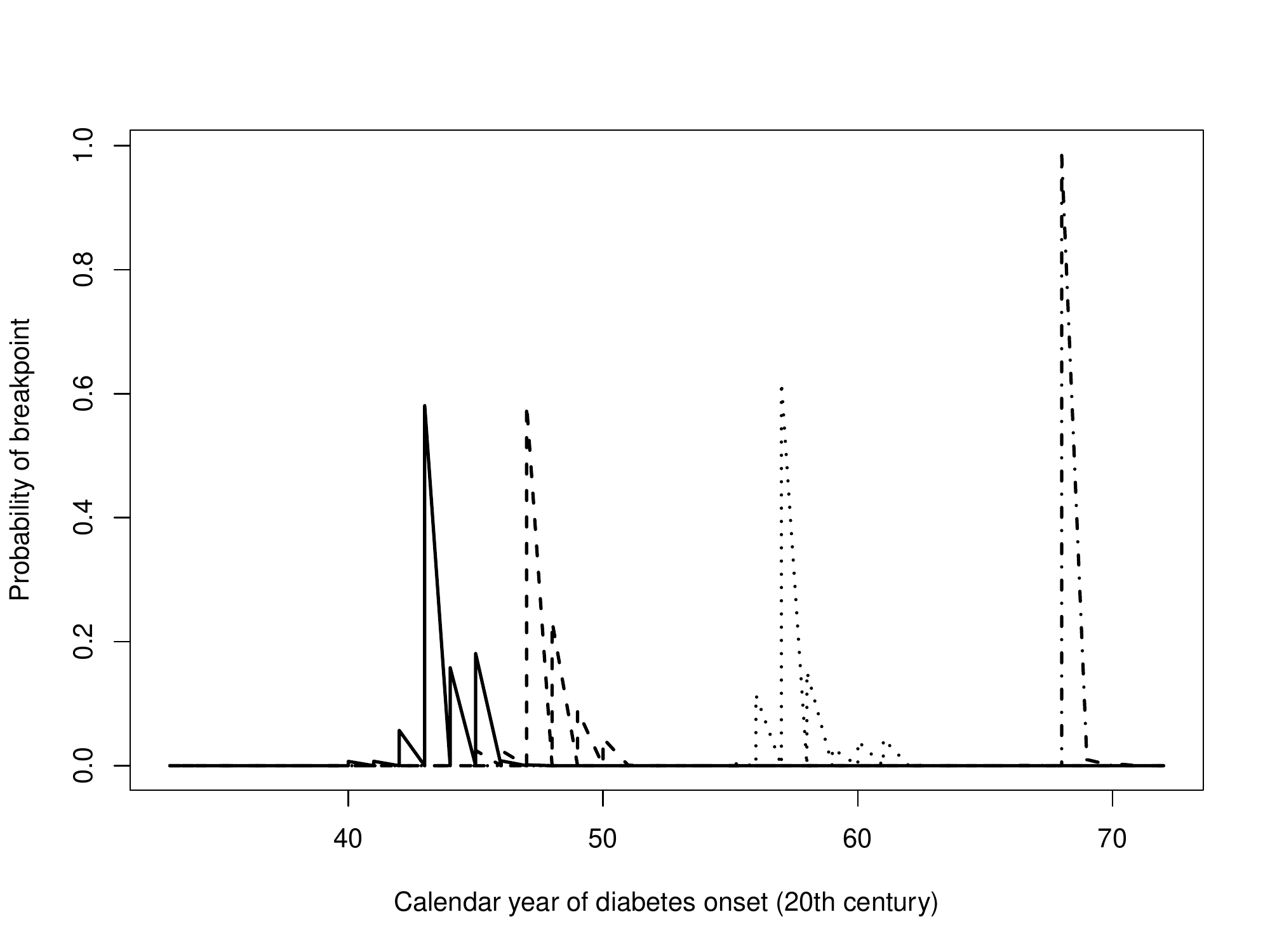}\\
\end{tabular}
\caption{Marginal distributions of the breakpoints in the models with one, two, three and four breakpoints. The maximum a posteriori for the breakpoints are respectively: top-left $1948$, top-right $1948$ and $1962$, bottom-left $1946$, $1957$ and $1962$, bottom-right $1944$, $1948$, $1958$ and $1969$.}
\label{bpplots}
\end{figure}

The maximum a posteriori of the breakpoints have been computed in Figure~\ref{bpplots}. For example, from the model with only one breakpoint it seems that the survival of diabetics patients was different for individuals born before the year $1948$ than for individuals born after $1947$ with a probability of having a breakpoint equal to $93\%$. For the two breakpoints model the probabilities a posteriori are also very sharp, with a probability of having a breakpoint at year $1948$ equal to $77\%$ and a second breakpoint at year $1962$ equal to $93\%$. For the three breakpoint models the probability a posteriori start to get slightly more widespread. The breakpoints occur in $1946$, $1957$ and $1962$ and their probabilities a posteriori are equal respectively to $81\%$, $32\%$ and $63\%$. Finally, in the four breakpoints model, the probabilities a posteriori of the breakpoints get very wide. They occur in $1944$, $1948$,  $1958$ and $1969$ with probabilities equal respectively to $58\%$, $58\%$, $62\%$ and $99\%$. From these plots we would tend to choose the two breakpoints model as the computed probabilities a posteriori are still very sharp compared to the three and four breakpoints model. This intuition is confirmed by the BIC criterion (see Table~\ref{paramBIC}) which clearly indicates that the two breakpoints model gives the best fit to the data compared to all the other models.

In Table~\ref{paramBIC}, parameter estimates for the Cox model with exponential baseline have also been computed with gender as a covariate. 
For the two breakpoint models, we also derived confidence intervals for the parameter estimates using a bootstrap procedure. We drew $200$ bootstrap samples and for each sample, new breakpoint locations along with the baseline values and regression parameters of each segment were computed. As a consequence, this procedure provides valid confidence intervals that take into account both the uncertainties into the breakpoint locations and into the parameter estimates. The baseline values are slightly decreasing with respect to the calendar year of diabetes onset in the sense that men and women diagnosed at a latter time have a smaller hazard of death than individuals diagnosed at a latter year. Their values along with their $95\%$ confidence intervals are respectively equal to $0.0226$ $[0.0198;0.0273]$, $0.0082$ $[0.0066;0.0123]$ and $0.0028$ $[0.0014;0.0048]$ on the respective segments  $1933-1947$,  $1948-1961$,  $1962-1972$. Looking at the effect of gender we see that this effect is positively associated to the hazard on the first two segments (so from $1933$ until $1961$) while its effect is no longer significant on the last segment. For better interpretation, we give here the hazard ratios between men and women (instead of the regression parameters as presented in Table~\ref{paramBIC}). On the respective segments  $1933-1947$,  $1948-1961$,  $1962-1972$, the hazard ratios for gender along with their $95\%$ confidence intervals are respectively equal to $1.2916$ $[1.0619;1.5453]$, $1.5970$ $[1.1185;2.0865]$ and $1.4426$ $[0.9046;3.3970]$.
 
%
%
%

\begin{figure}[!p]
\begin{tabular}{cc}
\includegraphics[width=0.5\textwidth]{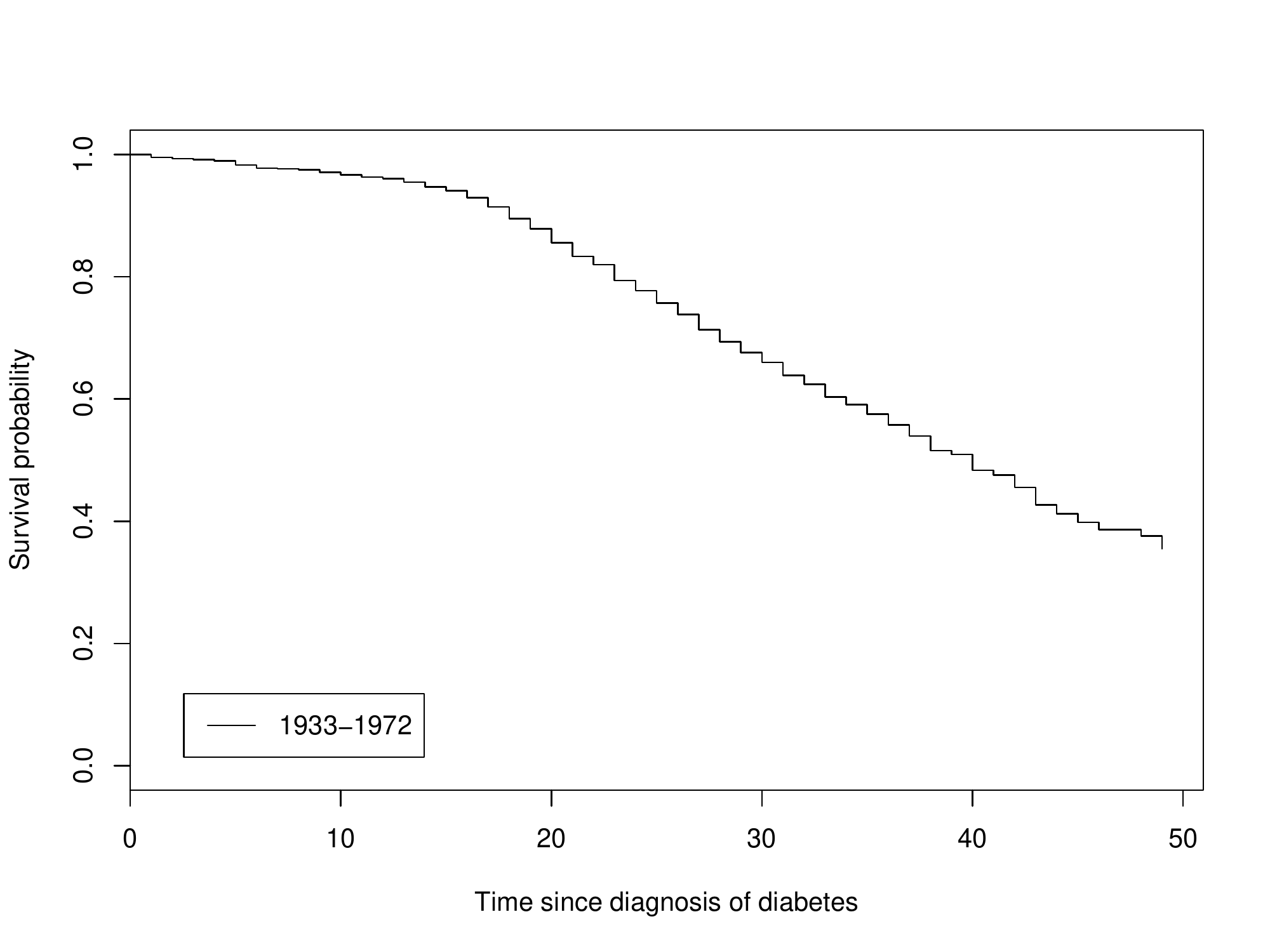}&\includegraphics[width=0.5\textwidth]{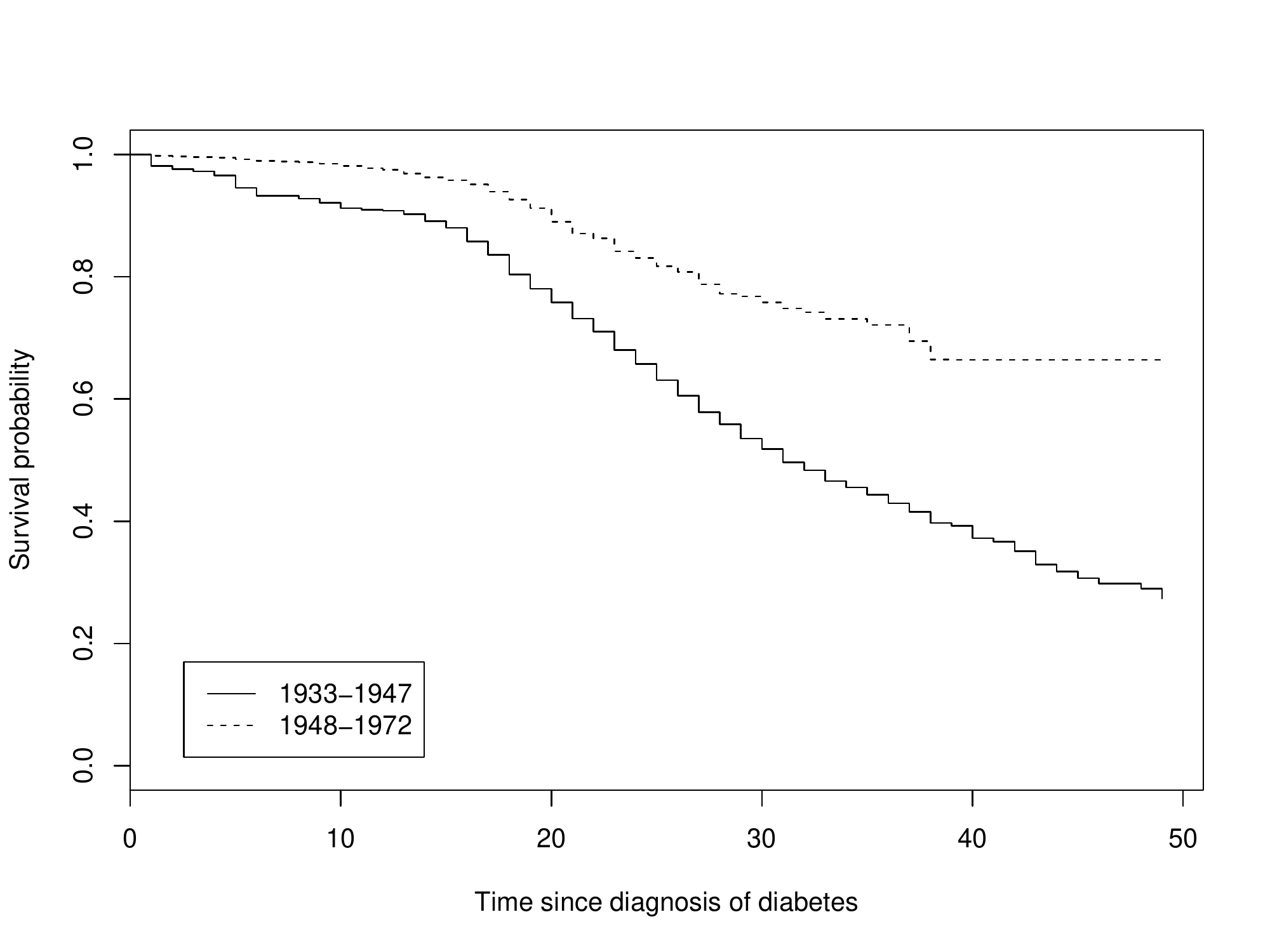}\\
\includegraphics[width=0.5\textwidth]{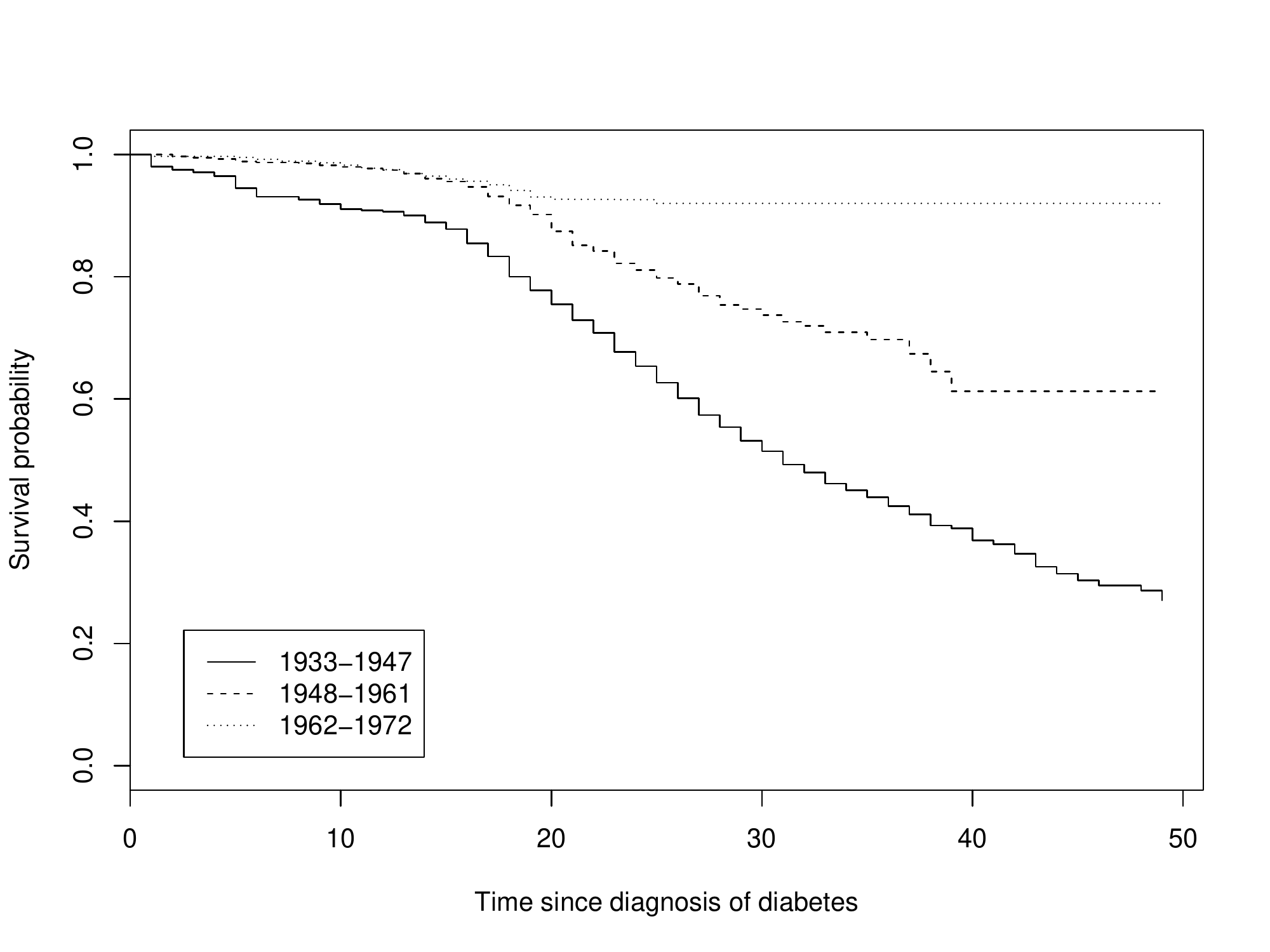}&\includegraphics[width=0.5\textwidth]{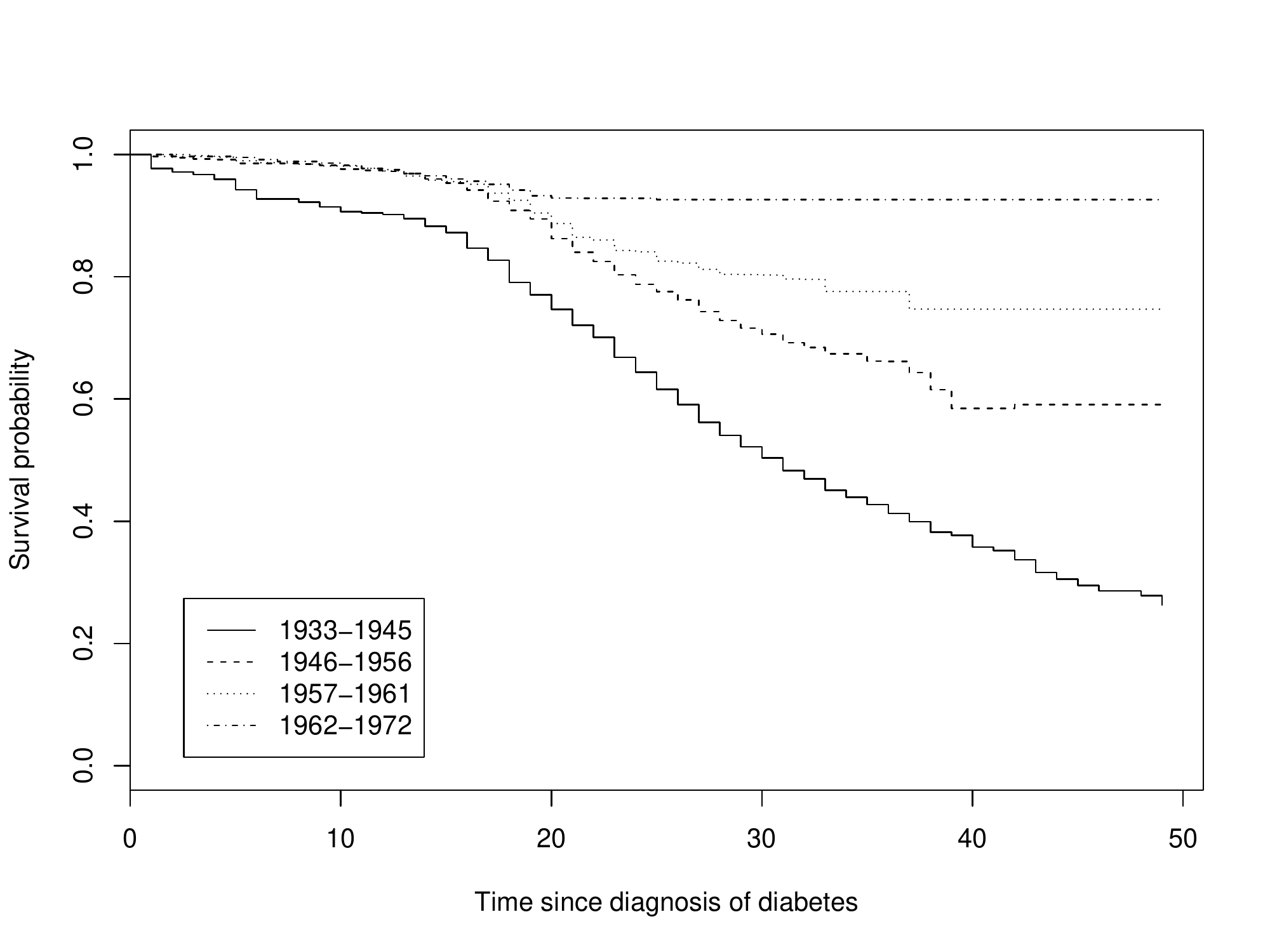}\\
\end{tabular}
\caption{Weighted Kaplan-Meier estimators in the models with zero (top-left), one (top-right), two (bottom-left) and three (bottom-right) breakpoints.}
\label{KMplots}
\end{figure}

Finally, nonparametric survival estimates have been computed using a weighted Kaplan-Meier estimator in Figure~\ref{KMplots}. The curves show a clear increase in the survival of patients according to the calendar year of diabetes onset. Patients diagnosed at a latter year have a greater survival than patients born at an earlier year. For example, in the two breakpoints model, the survival $30$ years after diagnoses of diabetes is equal to $51.4\%$, $73.8\%$, and $92\%$ for the respective diabetes of onset years  $1933-1947$, $1948-1961$ and $1962-1972$. Note that, using the bootstrap procedure as previously, one can also derive pointwise confidence intervals for these survival curves (not shown here).


The dataset has also been studied for the exponential model without adjusting by gender. The same breakpoints were found using the BIC criterion and the hazard and survival estimates were nearly identical.



\section{Discussion}\label{sec:conclusion}

In this article we introduced a new breakpoint model to detect heterogeneity in an ordered set of survival responses. In this model we suppose that abrupt changes can occur in the survival distribution of the event time. More specifically after specifying the number of segments, either the baseline hazard rates or the regression parameters are allowed to change in the different segments. Estimation in such a model is performed by an EM algorithm with use of constrained Hidden Markov Model (HMM) method as recently suggested by~\cite{Nuel}. The method proposes different specifications of the baseline and as shown by the simulation study, all different models provide both accurate estimates and accurate breakpoint locations. Interestingly, one can also obtain valid confidence intervals for quantities of interest such as the regression parameters or survival curves by taking into account both uncertainties in the location of the breakpoints and in the model parameters. This was illustrated on the Steno memorial hospital dataset through a bootstrap procedure. On this dataset the method was also shown to adapt to more realistic problematics such as left truncation. Taking into account ex-aequo individuals when ordered with respect to the calendar year of diabetes onset could also be achieved by correctly specifying the prior transition matrix. 
Clearly, the methods developed here could be readily extended to a more complex setting such as handling time dependent covariates or applying the method to recurrent events. Also, the methodology should be directly applicable to other survival models such as the Accelerated Failure Time Model (see~\citealp{kalbfleisch_book}; \citealp{Wei1992}) or the Aalen model (see~\citealp{Aalen1980};~\citealp{Scheike02}).

Strictly speaking our model only consider that abrupt changes may occur in terms of the survival distribution. This strong assumption clearly does not account for more continuous changes which is a classical drawback of breakpoint models. On the other hand, the resulting model is both parsimonious and highly interpretable since it provides a true segmentation of the original data. Also, one should note that slow changes in the hazard distribution could still be detected from our method: such data will result in a widely spread posterior probability distribution of the breakpoints.

As a measure of the fit of the breakpoint models to the data, a BIC criterion was derived for the parametric baseline models. This criterion turned out to be a very powerful tool since as shown in Section~\ref{sec:robust}, it seems to be very accurate to detect the correct number of breakpoints in a dataset. However note that no BIC criterion could be derived for the nonparametric baseline case. More generally it would be interesting to propose some kind of sequential testing procedure in order to find the number of breakpoints. In particular this will allow us to control the percentage of false discovery rate, that is the probability that more breakpoints than necessary are found in the dataset. This appears to be a complex problem and is left to future research work.

\section*{Acknowledgments}
The authors are very grateful to Professor Per Kragh Andersen for his valuable comments and for sharing with us the Steno memorial hospital dataset. 
This work is part of the DECURION project which was funded both by the IRESP and the french ``Ligue nationale contre le Cancer''. 
{\it Conflict of Interest}: None declared.

\section{Supporting Information}
%
Supporting Material is available online.

\bibliographystyle{biometrika}
\bibliography{biblio}

%

\begin{table}[!p]
\caption{Bias, variance, MSE of $\boldsymbol{\hat\beta}_{1}$, $\boldsymbol{\hat\beta}_{2}$, $\boldsymbol{\hat\beta}_{3}$ and estimations of the maximum probability of breakpoints, average breakpoint locations along with their $95\%$ empirical confidence intervals from Scenario 1 to 4 (top to bottom).
\label{Table1}}
\centering
\resizebox{1.0\columnwidth}{!}{
\begin{tabular}{lccccccccc}
\hline
\\
\multicolumn{10}{c}{\bf Scenario 1: exponential baselines}\\
\\
\hline
    & Bias of $\boldsymbol{\hat\beta}$ & Variance of $\boldsymbol{\hat\beta}$ & MSE of  $\boldsymbol{\hat\beta}$ & MAP of BP12 & Mean of BP12 & 95$\%$ CI of BP12 & MAP of BP23  &  Mean of BP23 &95$\%$ CI  of BP23 \\ 
\hline
Exponential & 0.002 & 0.006 & 0.006 & 0.411 & 1000 & $994$-$1006$ & 0.032 & 2120 & $1662$-$2974$ \\ 
   & -0.002 & 0.015 & 0.015 &  &  &  &  &  &  \\ 
   & -0.052 & 0.706 & 0.709 &  &  &  &  &  &  \\ 
  Weibull & 0.002 & 0.007 & 0.007 & 0.408 & 1000 & $994$-$1006$ & 0.043 & 2216 & $1740$-$2981$ \\ 
   & -0.002 & 0.011 & 0.011 &  &  &  &  &  &  \\ 
   & -0.007 & 0.407 & 0.407 &  &  &  &  &  &  \\ 
  Piecewise & 0.003 & 0.007 & 0.007 & 0.402 & 1000 & $994$-$1006$ & 0.069 & 2479 & $1800$-$2987$ \\ 
   & 0.000 & 0.009 & 0.009 &  &  &  &  &  &  \\ 
   & -0.066 & 0.574 & 0.578 &  &  &  &  &  &  \\ 
  Nonparametric & 0.002 & 0.007 & 0.007 & 0.429 & 1001 & $996$-$1007$ & 0.054 & 1954 & $1013$-$2995$ \\ 
   & -0.069 & 0.820 & 0.825 &  &  &  &  &  &  \\ 
   & -0.017 & 2.597 & 2.598 &  &  &  &  &  &  \\ 
\hline
\\
\multicolumn{10}{c}{\bf Scenario 2: Weibull baselines}\\
\\
\hline
\hline
    & Bias of $\boldsymbol{\hat\beta}$ & Variance of $\boldsymbol{\hat\beta}$ & MSE of  $\boldsymbol{\hat\beta}$ & MAP of BP12 & Mean of BP12 & 95$\%$ CI of BP12 & MAP of BP23  &  Mean of BP23 &95$\%$ CI  of BP23 \\ 
\hline
Exponential & -1.207 & 0.000 & 1.458 & 0.054 & 998 & $973$-$1016$ & 0.092 & 1943 & $1407$-$2002$ \\ 
   & 0.512 & 0.003 & 0.266 &  &  &  &  &  &  \\ 
   & 2.737 & 0.168 & 7.661 &  &  &  &  &  &  \\ 
  Weibull & -0.010 & 0.008 & 0.008 & 0.309 & 1002 & $996$-$1020$ & 0.154 & 1997 & $1978$-$2009$ \\ 
   & -0.009 & 0.008 & 0.008 &  &  &  &  &  &  \\ 
   & -0.043 & 0.255 & 0.257 &  &  &  &  &  &  \\ 
  Piecewise & -0.187 & 0.007 & 0.042 & 0.323 & 1001 & $995$-$1008$ & 0.192 & 1998 & $1983$-$2011$ \\ 
   & 0.031 & 0.007 & 0.008 &  &  &  &  &  &  \\ 
   & 0.007 & 0.304 & 0.304 &  &  &  &  &  &  \\ 
  Nonparametric & 0.000 & 0.010 & 0.010 & 0.332 & 1000 & $992$-$1008$ & 0.195 & 1998 & $1983$-$2012$ \\ 
   & -0.006 & 0.009 & 0.009 &  &  &  &  &  &  \\ 
   & -0.122 & 0.708 & 0.723 &  &  &  &  &  &  \\ 
\hline
\\
\multicolumn{10}{c}{\bf Scenario 3: piecewise constant baselines}\\
\\
\hline
    & Bias of $\boldsymbol{\hat\beta}$ & Variance of $\boldsymbol{\hat\beta}$ & MSE of  $\boldsymbol{\hat\beta}$ & MAP of BP12 & Mean of BP12 & 95$\%$ CI of BP12 & MAP of BP23  &  Mean of BP23 &95$\%$ CI  of BP23 \\ 
\hline
Exponential & -0.033 & 0.008 & 0.009 & 0.214 & 1001 & $986$-$1014$ & 0.043 & 1997 & $1854$-$2119$ \\ 
   & 0.002 & 0.010 & 0.010 &  &  &  &  &  &  \\ 
   & -0.007 & 0.016 & 0.016 &  &  &  &  &  &  \\ 
  Weibull & -0.013 & 0.007 & 0.008 & 0.216 & 1001 & $986$-$1014$ & 0.044 & 1994 & $1847$-$2111$ \\ 
   & 0.003 & 0.010 & 0.010 &  &  &  &  &  &  \\ 
   & -0.007 & 0.015 & 0.015 &  &  &  &  &  &  \\ 
  Piecewise & -0.007 & 0.008 & 0.008 & 0.217 & 1001 & $986$-$1014$ & 0.046 & 1990 & $1844$-$2116$ \\ 
   & 0.006 & 0.011 & 0.011 &  &  &  &  &  &  \\ 
   & -0.005 & 0.016 & 0.016 &  &  &  &  &  &  \\ 
  Nonparametric & 0.002 & 0.008 & 0.008 & 0.220 & 1002 & $991$-$1021$ & 0.042 & 1997 & $1847$-$2131$ \\ 
   & -0.001 & 0.010 & 0.010 &  &  &  &  &  &  \\ 
   & -0.006 & 0.015 & 0.015 &  &  &  &  &  &  \\
\hline
\\
\multicolumn{10}{c}{\bf Scenario 4: Gompertz baselines}\\
\\
\hline
    & Bias of $\boldsymbol{\hat\beta}$ & Variance of $\boldsymbol{\hat\beta}$ & MSE of  $\boldsymbol{\hat\beta}$ & MAP of BP12 & Mean of BP12 & 95$\%$ CI of BP12 & MAP of BP23  &  Mean of BP23 &95$\%$ CI  of BP23 \\ 
\hline
Exponential & -0.639 & 0.002 & 0.410 & 0.238 & 1000 & $992$-$1006$ & 0.027 & 1641 & $1015$-$2016$ \\ 
   & 0.196 & 0.020 & 0.058 &  &  &  &  &  &  \\ 
   & 0.575 & 0.035 & 0.366 &  &  &  &  &  &  \\ 
  Weibull & -0.212 & 0.005 & 0.050 & 0.352 & 1000 & $994$-$1006$ & 0.049 & 1994 & $1899$-$2079$ \\ 
   & 0.022 & 0.010 & 0.010 &  &  &  &  &  &  \\ 
   & 0.044 & 0.017 & 0.019 &  &  &  &  &  &  \\ 
  Piecewise & -0.076 & 0.007 & 0.013 & 0.378 & 1000 & $994$-$1006$ & 0.051 & 1989 & $1862$-$2080$ \\ 
   & 0.013 & 0.010 & 0.011 &  &  &  &  &  &  \\ 
   & 0.028 & 0.019 & 0.020 &  &  &  &  &  &  \\ 
  Nonparametric & 0.006 & 0.008 & 0.008 & 0.420 & 1000 & $991$-$1006$ & 0.049 & 2009 & $1928$-$2137$ \\ 
   & -0.004 & 0.011 & 0.011 &  &  &  &  &  &  \\ 
   & -0.023 & 0.165 & 0.165 &  &  &  &  &  &  \\ 
\noalign{\smallskip}\hline
\end{tabular}
}
\label{Table1}
\end{table}

\begin{table*}[!p]
\caption{Proportion of selected models using the AIC and BIC criterion for either the exponential baseline estimator or the piecewise constant hazard baseline estimator. Left side: when there is no breakpoints in the population. Right side: when the true number of breakpoints is two.}{
\begin{minipage}{.5\linewidth}
 \centering
\scalebox{0.85}{\begin{tabular}{c|cc|cc}
Number&\multicolumn{2}{c|}{Exponential estimator}&\multicolumn{2}{c}{Pch estimator}\\
 of bp&AIC&BIC&AIC&BIC\\
\hline\hline
{\bf 0}& 0.870&1&0.917&1\\
1&0.097&&0.066&\\
2& 0.024& & 0.015 &\\ 
  3 & 0.003&& 0.002 &\\ 
  4 & 0.003 &&  &\\ 
  5 &  &&  &\\ 
  6 & 0.003 && &\\ 
 \end{tabular}}
 \end{minipage}%
\begin{minipage}{.5\linewidth}
  \centering
\scalebox{0.85}{
 \begin{tabular}{c|cc|cc}
Number&\multicolumn{2}{c|}{Exponential estimator}&\multicolumn{2}{c}{Pch estimator}\\
 of bp&AIC&BIC&AIC&BIC\\
\hline\hline
0& &&&\\
1&&&&0.071\\
{\bf 2}& 0.801&0.987 & 0.872 &0.929\\ 
  3 & 0.116&0.013& 0.091 &\\ 
  4 & 0.047 && 0.025 &\\ 
  5 & 0.018 && 0.009 &\\ 
  6 & 0.018 && 0.003 &\\ 
 \end{tabular}\label{null} }
  \end{minipage}}
\end{table*}
%
%
%
%

\begin{table}[!p]
\caption{$\lambda$'s and $\beta$'s estimates in the Cox model adjusted by gender with exponential baseline for the models with zero, one, two, three and four breakpoints along with their BIC criterion.
}
\centering
\begin{tabular}{cccccc}
\hline\noalign{\smallskip}
 &No bp&  One bp& Two bp& Three bp& Four bp\\ 
  &&  1948& 1948, 62& 1946, 57, 62&  1944, 48, 58, 69\\ 
\noalign{\smallskip}\hline\noalign{\smallskip}
$\hat{\lambda}_1$ & 0.012& 0.022 & 0.023 & 0.023 & 0.024 \\ 
  $\hat{\lambda}_2$ && 0.006 & 0.008 & 0.011 & 0.015 \\ 
  $\hat{\lambda}_3$ & && 0.003 & 0.006 & 0.009 \\ 
  $\hat{\lambda}_4$ & & & & 0.003 & 0.004 \\ 
  $\hat{\lambda}_5$ & & & & & 0.001 \\ 
  $\hat{\beta}_1$ &0.278& 0.256 & 0.256 & 0.257 & 0.221 \\ 
  $\hat{\beta}_2$ & &0.477 & 0.468 & 0.344 & 0.357 \\ 
  $\hat{\beta}_3$ & & & 0.366 & 0.590 & 0.407 \\ 
  $\hat{\beta}_4$ & & &  & 0.377 & 0.509 \\ 
  $\hat{\beta}_5$ & & &  & & -0.101 \\ 
  BIC & 7426.405& 7214.413 & 7179.012 & 7187.442 & 7194.631 \\ 
\noalign{\smallskip}\hline
\end{tabular}
\label{paramBIC}
\end{table}


\end{document}